\documentclass[12pt]{article}

\usepackage{amsmath,amssymb,cite,axodraw,rotating,calc,graphics,epic,eepic,FeynMan,epsf}

\def\theequation{\arabic{section}.\arabic{equation}}

\newcommand {\ord} {\mathcal{O}}
\newcommand {\Lag} {\mathcal{L}}
\newcommand {\QCD} {\operatorname{QCD}}

\newcommand {\YMBKT} {\operatorname{5DQCD}}

\newcommand {\GF} {\operatorname{GF}}
\newcommand {\FP} {\operatorname{FP}}

\newcommand {\sss} {\scriptstyle}
\newcommand {\Zb} {\mathbb{Z}} 
\newcommand {\I} {\operatorname{I}}
\newcommand {\II} {\operatorname{II}}

\newcommand {\shift} {\hspace{3mm}}

\begin{document}

\begin{flushright}
WUE-ITP-2004-035\\
hep-ph/0411258\\
November 2004
\end{flushright}

\bigskip

\begin{center}
{\Large {\bf Quantization and High Energy Unitarity of}}\\[0.3cm]
{\Large {\bf 5D Orbifold Theories with Brane Kinetic Terms}}\\[1.4cm] 
{\large  Alexander M\"uck$^{\, a, b}$, Lars Nilse$^{\, c}$, Apostolos
Pilaftsis$^{\, c}$ and Reinhold R\"uckl$^{\,a}$}\\[0.4cm]
$^a${\em Institut f\"ur Theoretische Physik und Astrophysik,
         Universit\"at W\"urzburg,\\ Am Hubland, 97074 W\"urzburg, 
         Germany}\\[0.2cm]
$^b${\em School of Physics, University of Edinburgh,\\
         King's Buildings, Edinburgh EH9 3JZ, United Kingdom}\\[0.2cm]
$^c${\em School of Physics and Astronomy, University of Manchester,\\
         Manchester M13 9PL, United Kingdom}
\end{center}
\vskip1.cm   

\centerline{\bf ABSTRACT}   
\noindent
Five-dimensional field  theories compactified on an $S^1/\mathbb{Z}_2$
orbifold naturally  include local brane  kinetic terms at the orbifold
fixed points at the  tree as well as the  quantum level.  We study the
quantization of these  theories before the Kaluza--Klein reduction and
derive the  relevant Ward  and Slavnov--Taylor identities  that result
from the underlying gauge  and Becchi--Rouet--Stora symmetries of  the
theory.  With the help of these identities, we obtain a generalization
of the equivalence   theorem, where the known  high-energy equivalence
relation between the longitudinal  Kaluza--Klein gauge modes and their
respective would-be Goldstone bosons    is extended to    consistently
include   the   energetically  suppressed  terms  in   the high-energy
scatterings.   Demanding perturbative    unitarity,  we compute  upper
limits on the number  of the Kaluza--Klein modes.   We find that these
limits weakly depend on the size of the brane kinetic terms.


\newpage
\setcounter{equation}{0}
\section{Introduction}
\label{introduction}

Field  theories  that  realize  extra   compact  dimensions  offer new
perspectives~\cite{SS,Hosotani}  to address  problems  associated with
the breaking of  the electroweak gauge symmetry  of the Standard Model
(SM).   One  interesting aspect of  these  theories  is that the extra
components of the gauge fields may  acquire a vacuum expectation value
 through the  so-called Hosotani mechanism~\cite{Hosotani} and so
play the  role   of  the Higgs field~\cite{Quiros}.      Another novel
possibility for electroweak   symmetry breaking emerges  if  the gauge
symmetry is   broken   explicitly  by boundary  conditions,    e.g.~by
compactifying the theory on an interval~\cite{Csaki}.

Higher dimensional field theories, however, are not renormalizable, in
the sense that only a finite number of counterterms would be needed to
fix the  ultra-violet (UV) infinities  to all loop orders.  Therefore,
higher  dimensional  field theories should   be  regarded as effective
theories, and  as such,    they  can  play    a significant  role   in
understanding the low-energy  properties of a more fundamental  theory
that  could, for  example,  be of   stringy origin.  Their  consistent
formulation  within the  context of perturbation   theory should still
respect basic field-theoretic properties, such as gauge invariance and
perturbative unitarity.

The present  study  focuses  on five-dimensional  (5D)  field theories
compactified  on an $S^1/\mathbb{Z}_2$ orbifold.  Orbifolding provides
a viable mechanism to  introduce chiral  fermions after the  so-called
Kaluza--Klein (KK) reduction of 5D theories to 4 dimensions.  However,
a  consistent  description of   orbifold  field theories  requires the
inclusion of  brane kinetic  terms (BKTs)  which are  localized at the
orbifold fixed points~\cite{Georgi,Carena}.   These  new operators are
local  and need  be  included  as  counterterms at  the tree-level  to
renormalize the UV infinities of local operators of the same form that
are generated at the quantum level.

In    this    paper    we    extend   our    earlier    approach    to
quantization~\cite{MPR} to 5D orbifold theories that include BKTs.  An
important aspect of our approach is that the higher dimensional theory
is quantized, e.g.~within the framework of generalized $R_\xi$ gauges,
{\em before} the  KK reduction.  Such a gauge-fixing  procedure is not
trivial, since both the  gauge-fixing and ghost sectors should contain
terms  localized   at  the  orbifold  fixed  points.    After  the  KK
decomposition and  integration of  the extra dimension,  the resulting
effective  4D theory  coincides  with  the one  that  would have  been
obtained  if  the   KK  gauge  fields  had  been   quantized  mode  by
mode~\cite{PS} in the conventional $R_\xi$ gauge.

One of  the advantages of our 5D  quantization  formalism is that both
the   usual  gauge  and   the   so-called  Becchi--Rouet--Stora  (BRS)
transformations~\cite{BRS} take on a simple and finite form.  In close
analogy to  the ordinary case of  4D Quantum Chromodynamics (QCD), the
invariance of the Lagrangian under these transformations gives rise to
the    corresponding     5D    Ward    and     Slavnov--Taylor    (ST)
identities~\cite{Ward,ST}.  With the aid  of these identities, we  can
derive a generalization of the Equivalence Theorem~(ET)~\cite{CLT,LQT}
within  the  context of 5D   orbifold theories~\cite{Dicus} that takes
account of the presence of BKTs and the energetically suppressed terms
in high-energy  scatterings.    Such a generalization   is obtained by
following a line of argument  very analogous to  that for the ordinary
4D case~\cite{CG}  and has been assisted by  the introduction of a new
formalism  of functional differentiation   that preserves the orbifold
constraints on the 5D fields.

As  any perturbative framework   of  field theory, higher  dimensional
theories should  also satisfy perturbative unitarity so  as to be able
to make credible predictions for  physical observables, such as  cross
sections  and decay widths.   Requiring  the validity of  perturbative
unitarity for the  $s$-wave amplitude, we  compute upper limits on the
number of the   KK modes.  The  restoration of   high-energy unitarity
differs  in  our  models from  the  one  taking place in   5D orbifold
theories  without  BKTs~\cite{Dicus,Dominici}.    In  the  presence   of  BKTs,
unitarity is   achieved by  delicate cancellations  among   the entire
infinite  tower of  the  KK modes.  

We find, however, that  the upper limits established from perturbative
unitarity have a weak  dependence on the actual  size of the BKTs.  In
particular, large values  of BKTs do not  seem to screen the effect of
the bulk terms.   This weak dependence of the  unitarity limits on the
size of the BKTs may  be understood from  first principles as follows.
High-energy   unitarity strongly depends  on  the  UV structure of the
theory,  since at high     energies distances much smaller   than  the
compactification radius  of  the theory  are  probed.  Therefore, only
terms related to the bulk dynamics appear to be relevant when studying
perturbative    unitarity constraints.  Evidently, such considerations
will be of immediate phenomenological relevance when constraining more
realistic higher-dimensional  models  into which  the well-established
Standard Model (SM) can be embedded.

The structure of the paper is as follows.  In Section~2, after briefly
reviewing the  ordinary QCD case,   we study  the  quantization of  5D
orbifold  field  theories with  BKTs  before  the  KK reduction.    In
addition, we derive the  relevant Ward and  ST identities which are  a
consequence of the underlying gauge and  BRS symmetries of the theory.
With  the  help of   these identities,   we   derive in   Section~3  a
generalization of the ET, which we call the 5D generalized equivalence
theorem  (GET).  The 5D GET   provides a complete equivalence relation
between the longitudinal KK gauge  modes and their respective would-be
Goldstone bosons      that  takes   consistently  into    account  the
energetically suppressed terms  in high-energy scatterings.  Section~4
is devoted to the computation of upper limits on the  number of the KK
modes which   are obtained by  requiring the  validity of perturbative
unitarity.  We  also analyze the  dependence of these unitarity limits
on  the actual size  of the BKTs.  Technical  aspects of our study are
presented in  the   appendices.   Finally, Section~5    summarizes our
conclusions.


\setcounter{equation}{0}
\section{Ward and Slavnov--Taylor Identities}\label{ward}

We  will first  review the  basic  formalism of  quantization and  the
derivation of the  Ward and ST identities for  the conventional 4D QCD
case. This will  help us to set up our  conventions and obtain insight
into analogous  considerations concerning quantization  and derivation
of the corresponding  Ward and ST identities within  the context of 5D
orbifold field theories with~BKTs.

\subsection{4D QCD}
\label{wardQCD}

The ordinary 4D QCD Lagrangian quantized in the $R_\xi$ gauge is given
by
\begin{equation}
\label{LagQCD}
\Lag_{\QCD}\ =\ -\,\frac{1}{4} F^a_{\mu \nu} F^{a \, \mu \nu}\: +\:
\Lag_{\GF}\: +\: \Lag_{\FP}\,,
\end{equation}
where  $F^a_{\mu \nu}   =  \partial_{\mu} A^a_{\nu}  -  \partial_{\nu}
A^a_{\mu}+g f^{abc} A^b_{\mu} A^c_{\nu}$  is the field-strength tensor
of the     gluon  field  $A^a_{\mu}$.     Moreover, $\Lag_{\GF}$   and
$\Lag_{\FP}$ denote  the parts of the  Lagrangian, associated with the
gauge-fixing scheme and the Faddeev--Popov ghosts:
\begin{eqnarray}
\Lag_{\GF}  &=&  -\, \frac{1}{2  \xi} \big(F[A^a_{\mu}]\big)^2\ =\
-\,   \frac{1}{2 \xi}\;  (\partial^{\mu}    A^a_{\mu})^2\;,\nonumber\\
\Lag_{\FP} &=& \overline{c}^a\,     \frac{\delta  F[A^a_{\mu}]}{\delta
\theta^b} \,  c^b \ =\  \overline{c}^a\, \big(\delta^{ab} \partial^2\:
-\: g f^{abc}\, \partial^{\mu} A^c_{\mu}\,\big)\,c^b\; .
\end{eqnarray}
In the   absence  of  $\Lag_{\GF}$  and  $\Lag_{\FP}$,  the  remaining
(classical)  part of  the   QCD Lagrangian $\Lag_{\QCD}$ is  invariant
under the usual gauge transformations
\begin{align}
\label{4Dgaugetrans}
\delta A^a_{\mu}\ &= \ D^{ab}_{\mu}\, \theta^b\ =\ \big(\, \delta^{ab}
\partial_{\mu}\: -\: g f^{abc}\, A^c_{\mu}\,\big)\, \theta^b\; .
\end{align}
Hence, the gauge-invariant   part of the tree-level  effective action,
$\Gamma[A^a_{\mu}] = -\frac{1}{4}  \int d^4x \;  F^a_{\mu \nu} F^{a \,
\mu    \nu}$, satisfies the   relation  $\Gamma  [A^a_{\mu}] =  \Gamma
[A^a_\mu  +  \delta A^a_\mu]$,  from which  the following  master Ward
identity for QCD is easily derived:
\begin{equation}
\label{masterWardQCD}
\partial_{\mu} \frac{\delta \Gamma}{\delta A^a_{\mu}}\ -\
g f^{abc}\; \frac{\delta \Gamma}{\delta A^b_{\mu}} A^c_{\mu}\ =\ 0\; .
\end{equation}
There is a similarity of  the Ward identity~(\ref{masterWardQCD}) with
the  one  derived   with  the   background  field   method  (BFM)   in
QCD~\cite{Abbott}, which is  a consequence of  the gauge-invariance of
the effective   action with respect  to  gauge  transformations of the
background  fields. Therefore,  the  results  derived below from   the
classical part of $\Lag_{\QCD}$ will be valid within the BFM framework
as well.

After  functionally differentiating~(\ref{masterWardQCD}) with respect
to  gluon fields and subsequently setting  all the fields  to zero, we
obtain Ward identities  that relate  tree-level $n$-point correlation
functions  to each other. With   the convention that all momenta  flow
into the vertex, these Ward  identities may be graphically represented
as follows:
\begin{equation}
\begin{split}
-i k_{\mu} \BBB{a \; \mu}{b \; \nu}{c \; \rho}{\shift k}{p}{\shift q}
&= \ g f^{abd} \BB{d \; \nu}{c \; \rho}{\shift q}\: +\: g f^{acd} \BB{d \;
\rho}{b \; \nu}{\shift p}\\ &\phantom{}
\end{split}
\end{equation}
\begin{equation}
\begin{split}
-i k_{\mu} \BBBB{a \; \mu}{b \; \nu}{c \; \rho}{d \; \sigma}{k}{\shift
p}{q}{\shift r} &=\  g f^{abe} \BBB{e \; \nu}{c \; \rho}{d \;
\sigma}{k+p}{q}{\shift r}\: +\ g f^{ace} \BBB{e \;
\rho}{b \; \nu}{d \; \sigma}{k+q}{p}{\shift r}\\ &\phantom{}\\ 
&+\ g f^{ade} \BBB{e \; \sigma}{b \; \nu}{c \; \rho}{k+r}{p}{\shift q}\\ 
&\phantom{}\\ &\phantom{}
\end{split}
\end{equation}

We  now   turn  our   attention   to  the    complete  QCD  Lagrangian
(\ref{LagQCD}) quantized in the covariant $R_\xi$~gauges. Although the
gauge-fixing    and  ghost   terms break   the   gauge  invariance  of
$\Lag_{\QCD}$,  the  complete  QCD  Lagrangian $\Lag_{\QCD}$  is still
invariant under the so-called BRS transformations~\cite{BRS}:
\begin{eqnarray}
\label{BRSQCD}
s A^a_{\mu} &=& \ D^{ab}_{\mu} c^b\;,\nonumber\\
s\, c^a  &=& -\; \frac{g f^{abc}}{2} c^b c^c\;,\\
s\, \overline{c}^a &=& \frac{F[A^a_{\mu}]}{\xi}\ =\ 
\frac{\partial^{\mu} A^a_{\mu}}{\xi}\ .\nonumber
\end{eqnarray}
Note that the above form of BRS transformations is nilpotent after the
anti-ghost equation  of motion is  imposed. The nilpotency  of the BRS
transformations is an important property that ensures unitarity of the
$S$-matrix  operator     in     the   subspace    of   the    physical
fields~\cite{Pokorski,SMreviews}.

Following the  standard path-integral quantization approach, we define
the generating functional $Z$ of the connected Green functions through
the relation
\begin{equation}
e^{iZ}\, = \, \int\! DA \, Dc \, D\bar{c} \, \exp \Big[\, 
i\int\! d^4x\, \big(\, \Lag_{\QCD} + J^{a \mu} A^a_{\mu} +
\overline{D}^a c^a + \overline{c}^a D^a + K^{a \mu} s A^a_{\mu} + M^a
sc^a\, \big)\,\Big]\ .
\end{equation}
In the above, $J^{a \mu}$,  $\overline{D}^a$ and $D^a$ are the sources
for the  gluons, ghosts and  anti-ghosts,  respectively. As usual,  we
have included the additional sources $K^{a \mu}$ and $M^a$ for the BRS
variations $s  A^a_{\mu}$  and $s   c^a$ that  are   non-linear in the
fields.

Given  the  invariance   of  the  path-integral    measure   under BRS
transformations, it is not difficult to show that
\begin{equation}
  \label{Zgen}
Z \big[\, J^{a \mu},\ \overline{D}^a,\ D^a,\ K^{a \mu},\  M^a\,\big] 
\ =\ 
Z \big[\,J^{a \mu} - \mbox{$\frac{\omega}{\xi}$}\,\partial^\mu D^a,\
\overline{D}^a,\ D^a,\ K^{a \mu} + \omega J^{a \mu},\ 
M^a - \omega \overline{D}^a\,\big]\, ,
\end{equation}
where  $\omega$ is  a  Grassmann parameter,  i.e.~$\omega^2  =  0$. An
immediate  consequence of~(\ref{Zgen}) is   the master ST identity for
QCD
\begin{equation}
 \label{4DST}
J^{a \mu} \frac{\delta Z}{\delta K^{a \mu}}\ -\ \overline{D}^a
\frac{\delta Z}{\delta M^a}\ +\ \frac{1}{\xi}\, D^a\, \partial^{\mu}
\frac{\delta Z}{\delta J^{a \mu}}\ =\ 0\, .
\end{equation}
As before, a number of ST identities can be derived from~(\ref{4DST}),
using functional differentiation  techniques. 

An  alternative and  perhaps  more practical approach  to  deriving ST
identities is to use the BRS invariance of  the Green function defined
by  means of the time-ordered  operator formalism. To give an example,
let us  consider the BRS  invariance of  the Green function associated
with the fields $\bar{c}^a(x) A^b_{\nu}(y) A^c_{\rho}(z)$, i.e.
\begin{equation}
  \label{ersteZeile}
s\, \langle 0 | T \overline{c}^a(x) A^b_{\nu}(y) A^c_{\rho}(z) | 0
\rangle\ = \ 0\; .
\end{equation}
Employing  the   BRS transformations given   by~(\ref{BRSQCD}),  it is
straightforward to obtain the expression
\begin{eqnarray}
  \label{zweiteZeile} 
\frac{1}{\xi}\, \partial^{\mu}_{\sss x}\,
\langle 0 | T A^a_{\mu} (x) A^b_{\nu} (y) A^c_{\rho} (z) | 0 \rangle\:
-\: \partial^{\sss y}_{\nu} \langle 0 | T \overline{c}^a(x) c^b(y)
A^c_{\rho}(z)| 0 \rangle\
-\: \partial^{\sss z}_{\rho} \langle 0 | T
\overline{c}^a(x) A^b_{\nu} (y) c^c(z) | 0 \rangle\: &&\nonumber\\ 
+\:  g f^{bde} \langle 0 | T \overline{c}^a(x) c^d(y) A^e_{\nu} (y)
A^c_{\rho}(z) | 0 \rangle\: 
+\: g f^{cde} \langle 0 | T \overline{c}^a(x)
A^b_{\nu}(y) c^d(z) A^e_{\rho}(z) | 0 \rangle\ =\ 0\, .&&\nonumber\\
\end{eqnarray}
The last  two terms   on  the  LHS of~(\ref{zweiteZeile})   depend  on
bilinear   fields  evaluated at   the  same  space-time   point $y$ or
$z$. These terms do not have one-particle poles for the external lines
attached to those points, and therefore cancel on-shell. The resulting
on-shell ST identity for the involved one-particle irreducible Green 
functions may then be represented graphically as follows:
\begin{equation}
p^{\mu} \; \BBBblob{a \; \mu}{b \; \nu}{c \; \rho}{p}{q}{\shift r} -\
q_{\nu} \GGBblob{a}{b}{c \; \rho}{p}{q}{\shift r} -\ r_{\rho}
\GBGblob{a}{b \; \nu}{c}{p}{q}{\shift r} \ =\ 0 \; ,
\end{equation}

\vspace{1.cm}
\noindent
with $q^2 =  r^2 = 0$.  It  is not difficult to check  the validity of
the   above   on-shell   ST    identity   for   the   tree-level   QCD
Lagrangian~(\ref{LagQCD}).

For  our purpose  of   studying  high-energy  unitarity  in $2\to   2$
scatterings, more  useful  are ST identities  that  apply  to on-shell
4-point functions, e.g.~to the   product  of fields:   $A^a_{\nu}  (x)
A^b_{\nu} (y) A^c_{\rho}  (z) A^d_{\sigma} (w)$.  Again, starting from
the identity
\begin{equation}
  \label{dritteZeile}
s\; \langle 0 | T \; \bar{c}^a(x) [\partial^{\nu} A^b_{\nu} (y)] 
[\partial^{\rho} A^c_{\rho} (z)] [\partial^{\sigma} A^d_{\sigma} (w)]
| 0 \rangle\ =\ 0\, ,
\end{equation}
we arrive at the ST identity which may be given diagrammatically by
\begin{equation}
  \label{vierteZeile}
k_{\mu} \, p_{\nu} \, q_{\rho} \, r_{\sigma} \; 
\BBBBblob{\mu \; a}{\nu \; b}{\rho \; c}{\sigma \; d}{k}{\shift
p}{q}{\shift r}\ =\ 0\ .
\end{equation}

\vspace{1.cm}
\noindent
As we will see in the next section and Section~3, it  is the 5D analog
of the above on-shell ST identity which lies at the heart of the proof
of the GET for the 5D orbifold field theories.


\subsection{5D Orbifold Theories with Brane Kinetic Terms}\label{wardCYMBKT}

We  will  now  study   a  5D  Yang--Mills   theory, such   as  5D QCD,
compactified on an  $S^1/\mathbb{Z}_2$ orbifold.  As has been observed
in~\cite{Georgi,Carena}, orbifold compactification introduces BKTs for
the 5D gluon field at the orbifold fixed points $y=0$ and $y = \pi R$.
These BKTs naturally emerge  as counterterms at the  tree level, so as
to cancel the  UV infinities of the  same  form that are generated  by
radiative corrections~\cite{Quiros}.   Here and  in the  following, we
denote Lorentz indices enumerating the 5 dimensions with capital Roman
letters $M,N$ etc., with $M = \mu, 5$ ($\mu = 0,1,2,3$), and the fifth
compact dimension with $y\equiv x^5 \in (-\pi R,\pi R]$.

To  avoid excessive complication in  our analytic results, we consider
5D QCD compactified on an $S^1/\mathbb{Z}_2$  orbifold with one BKT at
$y=0$. However, our discussion can very analogously  carry over to the
general case with two BKTs at $y=0$ and $y =  \pi R$.  To be specific,
the 5D QCD Lagrangian of interest reads
\begin{equation}
\label{LagYMBKT}
\Lag_{\YMBKT} (x,y)\ =\ 
-\, \frac{1}{4}\, \Big( 1\: +\: r_c\,\delta(y) \Big)\,  
F^a_{MN} F^{a \; MN}\ +\ \Lag_{\rm 5D\GF}\ +\ \Lag_{\rm 5D\FP}\,,
\end{equation}
where $F^a_{MN}  = \partial_M A^a_N  - \partial_N A^a_M +  g_5 f^{abc}
A^b_M A^c_N$ is the corresponding field-strength tensor for the 5D QCD
gluon  $A^a_M$, and  $r_c$  is  taken to  be  a positive  dimensionful
coupling  for the  BKT  at $y=0$.\footnote{If  $r_c$  is negative  and
$|r_c| \leq 2\pi R$, the theory will contain undesirable negative norm
states~\cite{Santiago}.}   In addition,  the terms  $\Lag_{\rm 5D\GF}$
and $\Lag_{\rm 5D\FP}$ describe the 5D gauge-fixing and Faddeev--Popov
ghost sectors to be discussed in detail below.

In  order that  the 5D orbifold  theory  includes ordinary QCD with  a
massless  gluon, it is  sufficient for the  5D  gluon field $A^a_M$ to
obey the following constraints:
\begin{equation}
  \label{cond}
A^a_M(x,y)\ =\ A^a_M(x, y+2 \pi R)\,,\quad 
A^a_{\mu}(x,y)\ =\ A^a_{\mu}(x,-y)\,,\quad 
A^a_5(x,y)\ =\ - A^a_5(x,-y)\; .
\end{equation}
Observe  that $A^a_\mu$   ($A^a_5$)    is   even  (odd)   under    the
$\mathbb{Z}_2$ reflection,  $y \to  -  y$. Although  the BKT  at $y=0$
breaks explicitly  the  higher-dimensional Lorentz invariance   of the
Lagrangian~(\ref{LagYMBKT}) down to 4  dimensions, the classical  part
of the  5D  QCD  Lagrangian  remains invariant   under the  5D   gauge
transformations
\begin{equation}
\label{5DgaugeTransformation}
\delta A^a_M\ =\ D^{ab}_M\, \theta^b \ = \
\big(\, \delta^{ab} \partial_M\: -\: g_5 f^{abc} A^c_M\,\big)\, \theta^b\, .
\end{equation}
For consistency,  the  5D gauge  parameter $\theta^a (x,y)$  has to be
periodic and even under $S^1/\mathbb{Z}_2$: $\theta^a (x,y) = \theta^a
(x, y+2 \pi R)$ and $\theta^a (x,y) = \theta^a (x,-y)$.

Given the  $S^1/\mathbb{Z}_2$ parities of   the 5D gluon  field $A^a_M
(x,y)$ and the gauge parameter $\theta^a  (x,y)$, these quantities can
be expanded in infinite series in  terms of orthonormal functions $f_n
(y)$ and $g_n  (y)$ that are even and  odd in $y$, respectively.  More
explicitly, we have
\begin{eqnarray}
\label{AmuExpansion}
A^a_\mu (x,y) &=& \sum_{n=0}^{\infty} A^a_{(n) \mu}(x)\, f_n(y)\;,\nonumber\\
A^a_5 (x,y) &=& \sum_{n=1}^{\infty} A^a_{(n) 5}(x)\, g_n(y)\;,\\
\theta^a(x,y) &=& \sum_{n=0}^{\infty} \theta^a_{(n)}(x)\, f_n(y)\; .\nonumber
\end{eqnarray}
An important property of the orthonormal  functions $f_n (y)$ and $g_n
(y)$    is that  the resulting  coefficients    $A^a_{(n) \mu}(x)$ and
$A^a_{(n) 5} (x)$, the so-called KK modes, are mass eigenstates of the
effective  theory, after  integrating  out the  extra  dimension.  For
example,  in the  limit of a  vanishing  BKT term,  $r_c   \to 0$, one
recovers the standard Fourier series   expansion, where the  functions
$f_n (y)$  and $g_n (y)$ are proportional  to $\cos  (ny/R)$ and $\sin
(ny/R )$, respectively.

In Appendix~\ref{masseigenmodeexpansion} we  derive the analytic forms
of the orthonormal functions $f_n  (y)$ and $g_n (y)$ for the complete
5D  orbifold theory  with BKTs.   To properly  deal with  the singular
behaviour of  the BKTs, e.g.~at  $y=0$, we introduce  a regularization
method to analytically define $f_n  (y)$ and $g_n (y)$ in the interval
$(-\epsilon,\epsilon)$, where $\epsilon$  is an infinitesimal positive
constant which  is taken to  zero at the  end of the  calculation.  In
their definition  interval $(-\pi R +\epsilon, \pi  R +\epsilon]$, the
$\epsilon$-regularized orthonormal functions are given by
\begin{eqnarray}
\label{fn}
f_n(y) &=& \frac{N_n}{\sqrt{2^{\delta_{n,0}} \pi R}} 
\times \begin{cases}
\ \cos m_n y\: -\: \frac{1}{2}\,m_n\,r_c\,\sin m_n y\,,\  
&\textrm{for}\ -\pi R + \epsilon <
y \leq -\epsilon\\
\ \cos m_n y\,,\ &\textrm{for} \ 
-\epsilon < y < \epsilon\\ 
\ \cos m_n y\: +\: \frac{1}{2}\,m_n\,r_c\,\sin m_n y\,,
&\textrm{for}\quad \epsilon \leq y 
\leq \pi R + \epsilon
\end{cases}\\[3mm]
  \label{gn}
g_n(y) &=& \frac{N_n}{\sqrt{2^{\delta_{n,0}} \pi R}} 
\times\begin{cases}
\ \sin m_n y\: +\: \frac{1}{2}\,m_n\,r_c\,\cos m_n y\,,\  &
\textrm{for}\ -\pi R+\epsilon <
y\leq -\epsilon \\
\ \sin m_n y\,,\ &\textrm{for}\  
-\epsilon < y < \epsilon\\ 
\ \sin m_n y\: -\: \frac{1}{2}\,m_n\,r_c\,\cos m_n y\,,\  &
\textrm{for}\quad \epsilon \leq y
\leq \pi R +\epsilon\; ,
\end{cases}
\end{eqnarray}
where 
\begin{equation}
  \label{Nn}
N_n^{-2}\ =\ 1 + \tilde{r}_c + \pi^2 R^2 \tilde{r}_c^2 m_n^2\;, 
\end{equation}
with  $\tilde{r}_c  =  r_c/(2  \pi   R)  \geq  0$.   Notice  that  the
orthonormal functions  $f_n (y)$  and $g_n (y)$  are gauge-independent
and related to each other through
\begin{equation}
  \label{partial}
\partial_5 f_n \ =\ -\, m_n g_n\,,\qquad \partial_5 g_n \ =\ m_n f_n\; ,
\end{equation}
where the derivatives  are understood to  be piecewise  defined within
the   respective sub-intervals  given   in~(\ref{fn}) and  (\ref{gn}).
Finally,  the   physical  masses    $m_n$  of  the  KK  gauge    modes
$A^a_{(n)\,\mu}$  are     obtained by    solving    numerically    the
transcendental equation~\cite{Carena}:
\begin{equation}
  \label{spectrum}
\frac{m_n r_c}{2}\ =\ -\, \tan \big(m_n \pi R\big)\; .
\end{equation}
In addition to the massless solution $m_0 = 0$, in the limits $r_c \to
0$  and  $r_c \to \infty$, (\ref{spectrum})   has  the simple analytic
solutions $m_{n\ge 1} = n/R$ and $m_{n\ge 1} = (2n-1)/2R$, respectively.

Employing Fourier-like   convolution techniques developed  in Appendix
C.1, we   can calculate  the  Feynman  rules of  the  effective 4D  KK
theory. These   are listed in Appendix~A.    For example, applying the
convolution         techniques        to        the    5D        gauge
transformations~(\ref{5DgaugeTransformation}),  we find the  effective
gauge transformations for the KK modes
\begin{eqnarray}
\label{transBKT}
\delta A^a_{(n) \mu} &=& \partial_{\mu} \theta^a_{(n)}\: -\: 
g f^{abc} \sum_{m, l=0}^{\infty} \sqrt{2}^{\;
-1-\delta_{n,0}-\delta_{m,0}-\delta_{l,0}} \; \theta^b_{(m)}\, A^c_{(l)
\mu}\, \Delta_{n,l,m}\; ,\nonumber\\
\delta A^a_{(n) 5} &=& -\, m_n \theta^a_{(n)}\: -\: 
g f^{abc} \sum_{m=0, \; l=1}^{\infty} 
\sqrt{2}^{\; -1-\delta_{m,0}} \; \theta^b_{(m)}\, A^c_{(l) 5}\, 
\tilde{\Delta}_{n,l,m}\; ,
\end{eqnarray}
where $g  = g_5/\sqrt{2\pi R}$ is  a  dimensionless coupling constant.
In the  limit $r_c \to  0$, these transformations   reduce to the ones
stated    in~\cite{MPR}  without   BKTs,   with  the  identifications:
$\Delta_{n,l,m}    = \delta_{n,l,m}$    and $\tilde{\Delta}_{n,l,m}  =
\tilde{\delta}_{n,l,m}$.  The  definitions  of  $\Delta_{n,l,m}$   and
$\tilde{\Delta}_{n,l,m}$ are given in Appendix C.2.

A technical drawback  of~(\ref{transBKT})  is that  the number of  the
effective  gauge-field transformations  becomes infinite.  Our task of
deriving  Ward  and ST  identities will be  greatly facilitated  if we
introduce a   formalism of functional   differentiation on  an  $S^1 /
\Zb_2$  orbifold.    To  this end,  we    need  first  to define   the
$\delta$-function relevant to our $S^1 /  \Zb_2$ theory with BKTs.  An
appropriate  construction  can   be    achieved by   exploiting    the
completeness relation
\begin{equation}
 \label{deltaExp}
\delta (y-y'; r_c)\ =\ \sum_{n=0}^{\infty}\,\big[\, f_n (y)\, f_n (y')\: +\:
g_n (y)\, g_n(y')\,\big]\, .
\end{equation}
In the limit $r_c \to 0$, the $\delta$-function goes over to the known
form:
\begin{equation}
\delta (y;r_c=0)\ \equiv\ \delta (y) \ =\ \sum_{n=0}^{\infty} \,
\frac{1}{2^{\delta_{n,0}}\,\pi R}\ \cos \bigg( \frac{ny}{R} \bigg)\; .
\end{equation}
For  any periodic test function   $h(y)$ defined on  $S^1$,  it can be
shown that the defining property of the $\delta$-function,
\begin{equation}
\label{deltaDef}
\int_{-\pi R}^{\pi R} dy\, [ 1\, +\, r_c \delta (y) ]\; 
h(y)\, \delta (y;r_c)\ =\ h(0)\, ,
\end{equation}
holds  true.  In fact, the  proof  of~(\ref{deltaDef}) is based on the
orthonormality of the mass eigenmode wavefunctions  $f_n (y)$ and $g_n
(y)$.

We   are now in a  position  to introduce  a functional differentiation
formalism  that respects  the   $S^1/\Zb_2$ properties of the   fields
$A^a_\mu (x,y)$   and   $A^a_5 (x,y)$.  To   be  specific,  functional
differentiation on $S^1/\Zb_2$ may be defined as follows:
\begin{eqnarray}
\label{fdDef}
\frac{\delta A^a_{\mu}(x_1,y_1)}{\delta A^b_{\nu}(x_2,y_2)} &=&
\frac{1}{2}\,g_\mu^\nu\,\delta^{ab}\, \Big[\, \delta(y_1-y_2; r_c)\ +\
\delta(y_1+y_2; r_c)\,\Big]\, \delta^{(4)}(x_1-x_2)\; ,\nonumber\\
\frac{\delta A^a_5(x_1,y_1)}{\delta A^b_5(x_2,y_2)} &=&
\frac{1}{2}\, \delta^{ab}\, \Big[\,\delta(y_1-y_2; r_c)\ -\ \delta(y_1+y_2;
r_c)\,\Big]\, \delta^{(4)}(x_1-x_2)\;.
\end{eqnarray}
Notice that  functional   differentiation  with respect  to   $A^b_\nu
(x_2,y_2)$ ($A^b_5 (x_2,y_2)$) is even  (odd) under $y_2 \to -y_2$, as
it should be.

Like in the ordinary QCD case, we may  now derive Ward identities that
result from  the 5D  gauge invariance  of  the classical part  of  the
tree-level  effective action: $\Gamma  [ A^a_M  ] = -\frac{1}{4}\,\int
d^4x \int_{-\pi R}^{\pi R} dy \, [1  + r_c \delta (y)]\, F^a_{MN} F^{a
\; MN}$. Making use of the newly introduced functional differentiation
formalism, we obtain the master Ward identity for 5D QCD
\begin{equation}
\label{masterWard5DYM}
\partial_M \, \frac{\delta \Gamma}{\delta A^a_M}\ -\ g_5 f^{abc}
\frac{\delta \Gamma}{\delta A^b_M} A^c_M\ =\ 0\; .
\end{equation}
In order  to translate the 5D master  Ward identity into the effective
4D  one,   it proves  useful   to define   the  decompositions  of the
functional derivatives $\delta / \delta A^a_{\mu} (x,y)$ and $\delta /
\delta A^a_5 (x,y)$ in terms   of the orthonormal functions $f_n  (y)$
and $g_n (y)$:
\begin{eqnarray}
  \label{dd}
\frac{\delta}{\delta A^a_{\mu}(x,y)} &=& \sum_{n=0}^{\infty} f_n(y) \;
\frac{\delta}{\delta A^a_{(n) \mu}(x)}\,,\nonumber\\ 
\frac{\delta}{\delta A^a_5 (x,y)} &=& 
\sum_{n=1}^{\infty} g_n(y) \; \frac{\delta}{\delta A^a_{(n) 5}(x)}\ .
\end{eqnarray}
Observe  that the  definitions in~(\ref{dd})  are consistent  with the
functional differentiation rules stated in~(\ref{fdDef}). Substituting
(\ref{dd}) into (\ref{masterWard5DYM}) and integrating over $y$ yields
the effective master Ward identity for the compactified theory
\begin{equation}
  \label{masterWardCYM}
\begin{split}
\partial_{\mu} \frac{\delta \Gamma}{\delta A^a_{(n) \mu}}\ +\ m_n
\frac{\delta \Gamma}{\delta A^a_{(n) 5}}\ =\ \; &g f^{abc} \,
\sum_{m,l=0}^{\infty} \sqrt{2}^{\; -1-\delta_{n,0} - \delta_{m,0} -
\delta_{l,0}}\\ &\times\, \bigg(\, \frac{\delta \Gamma}{\delta A^b_{(m)
\mu}} A^c_{(l) \mu} \Delta_{m, n, l}\ +\ \frac{\delta \Gamma}{\delta
A^b_{(m) 5}} A^c_{(l) 5} \tilde{\Delta}_{m,n,l}\, \bigg)\; .
\end{split}
\end{equation}

By functional differentiation with respect to KK fields, we may derive
Ward identities among different $n$-point correlation functions, which
may graphically be represented as follows:
\begin{equation}
\label{wardCYM1}
\begin{split}
-i k_{\mu} \BBB{a \; \mu}{b \; \nu}{c \; \rho}{(n) \; k}{(m) \; p}{(l)
\; q} = &\ \  -m_n \SBB{a}{b \; \nu}{c \; \rho}{(n) \; k}{(m) \; p}{(l)
\; q} \\ &\phantom{} \\ &\phantom{} \\ +\ \sqrt{2}^{\; -1-\delta_{n,0}
- \delta_{m,0} - \delta_{l,0}} \, &\Delta_{n, m, l} \: g \big[ f^{abd}
\BB{d \; \nu}{c \; \rho}{(l) \; q} +\ f^{acd} \BB{d \; \rho}{b \;
\nu}{(m) \; p} \big]
\end{split}
\end{equation}

\bigskip

\begin{equation}
\begin{split}
-i k^{\mu} &\BBS{a \; \mu}{b \; \nu}{c}{(n) \; k}{(m) \; p}{(l) \; q} =\
-m_n \SBS{a}{b \; \nu}{c}{(n) \; k}{(m) \; p}{(l) \; q}\\ &\phantom{}
\\ &\phantom{}\\ &+\ \sqrt{2}^{\; -1-\delta_{n,0} - \delta_{m,0}} \, g
\, \big[ \Delta_{m,n,l} f^{abd} \BS{d \; \nu}{c}{(l) \; q} -\
\tilde{\Delta}_{m,n,l} f^{acd} \BS{b \; \nu}{d}{(m) \; p} \big]
\end{split}
\end{equation}

\bigskip

\begin{equation}
\begin{split}
-i k_{\mu} \BSS{a \; \mu}{b}{c}{(n) \; k}{(m) \; p}{(l) \; q} = &\ 
\sqrt{2}^{\; -1-\delta_{n,0}} \, \tilde{\Delta}_{m,n,l} \: g \, \big[
f^{abd} \SSLL{ d }{ c }{(l) \; q} +\ f^{acd} \SSLL{d}{b}{(m) \; p} \big]\\
&\phantom{}
\end{split}
\end{equation}

\bigskip

\begin{equation}
\begin{split}
-i k^{\mu} \BBBB{a \; \mu}{b \; \nu}{c \; \rho}{d \; \sigma}{(n) \;
k}{(m) \; p}{(l) \; q}{(k) \; r} \hspace{-3mm} = &\ \sum_{j=0}^{\infty}
g \, \sqrt{2}^{\; -1-\delta_{n,0} - \delta_{j,0}} \big[ \sqrt{2}^{\; -
\delta_{m,0}} \, \Delta_{m,n,j} f^{abe} \BBB{e \; \nu}{c \; \rho}{d \;
\sigma}{(j) \; k+p}{(l) \; q}{(k) \; r}\\ &\phantom{} \\ +
\sqrt{2}^{\; - \delta_{l,0}} \, \Delta_{l,n,j} &f^{ace} \BBB{e \;
\rho}{b \; \nu}{d \; \sigma}{(j) \; k+q}{(m) \; p}{(k) \; r}
\hspace{-5mm} +\ \sqrt{2}^{\; - \delta_{k,0}} \, \Delta_{k,n,j} f^{ade}
\BBB{e \; \sigma}{b \; \nu}{c \; \rho}{(j) \; k+r}{(m) \; p}{(l) \; q}
\hspace{-5mm} \big]\\ &\phantom{} \\ &\phantom{}
\end{split}
\end{equation}

\bigskip

\begin{equation}
\label{wardCYM5}
\begin{split}
-i k^{\mu} \BBSS{a \; \mu}{b \; \nu}{c}{d}{(n) \; k}{(m) \; p}{(l) \;
q}{(k) \; r} \hspace{-3mm} = &\ \sum_{j=0}^{\infty} g \, \sqrt{2}^{\;
-1-\delta_{n,0} - \delta_{j,0}} \big[ \sqrt{2}^{\; - \delta_{m,0}} \,
\Delta_{m,n,j} f^{abe} \BSS{e \; \nu}{c}{d}{(j) \; k+p}{(l) \; q}{(k)
\; r}\\ &\phantom{} \\ +\ \tilde{\Delta}_{l,n,j} &f^{ace} \SBS{e}{b \;
\nu}{d}{(j) \; k+q}{(m) \; p}{(k) \; r} \hspace{-5mm} +\
\tilde{\Delta}_{k,n,j} f^{ade} \SBS{e}{b \; \nu}{c}{(j) \; k+r}{(m) \;
p}{(l) \; q} \hspace{-5mm} \big]\\ &\phantom{}\\ &\phantom{}
\end{split}
\end{equation}
At  this point,  we should  remark that  $A^a_{(n) 5}$  satisfies Ward
identities very  analogous to those  found for the  would-be Goldstone
bosons in  spontaneously broken gauge  theories~\cite{Papa}. Moreover,
we    should     stress    again    that    the     above    5    Ward
identities~(\ref{wardCYM1})--(\ref{wardCYM5})  will also  hold  in the
case of a BFM formulation of 5D Yang--Mills theories~\cite{Buras}.

We will now discuss the part of the 5D Lagrangian in~(\ref{LagYMBKT}),
associated with the  gauge-fixing scheme and  the ghost sector. Within
the   framework of the generalized  $R_\xi$   gauges, all mixing terms
between  the KK   gauge fields  $A^a_{(n)\,\mu}$  and their   would-be
Goldstone  bosons $A^a_{(n)\,   5}$ are absent.   To incorporate  this
property, we proceed very analogously to~\cite{MPR}  and choose the 5D
gauge-fixing functional
\begin{equation}
 \label{gfFunctional} 
F[A^a_M]\ =\ \partial^{\mu} A^a_{\mu}\: -\: \xi\, \partial_5 A^a_5\; .
\end{equation}
However, we multiply the 5D  gauge-fixing and the Faddeev--Popov ghost
terms with  the  expression $[ 1  +  r_c \delta  (y)]$  which includes
localized   interactions of the  same  form  as those  present in  the
gauge-kinetic part of the Lagrangian, i.e.
\begin{eqnarray}
  \label{gfTerm}
\Lag_{\rm 5D\GF}(x,y) &=& -\, \big[  1 +   r_c  \delta
(y) \big]\, \frac{1}{2 \xi}\; 
\big(F[A^a_M]\big)^2\ =\ -\, \big[  1 +   r_c  \delta
(y) \big]\, \frac{1}{2 \xi}\,  \Big(\,
\partial^{\mu} A^a_{\mu}\: -\: \xi\, \partial_5
A^a_5\,\Big)^2\;,\nonumber\\[3mm] 
\Lag_{\rm 5D\FP}(x,y) &=& \big[  1 +   r_c  \delta
(y) \big]\, \bar{c}^a\; \frac{\delta F[A^a_M]}{\delta
\theta^b}\; c^b\\
&=& \big[  1 +   r_c  \delta
(y) \big]\,\bar{c}^a\, \Big[\, \delta^{ab}\, \Big( \partial^2\, -\,
\xi\, \partial_5^2\Big)\: -\: g_5 f^{abc}\,
\Big(\,\partial^{\mu} A^c_{\mu}\, -\, \xi\, \partial_5\,
A^c_5\,\Big)\, \Big] c^b\; ,\nonumber
\end{eqnarray}
where $c^a (x,y)$ and $\bar{c}^a  (x,y)$  are the  5D ghosts that  are
even under $\Zb_2$. 

An      important   constraint     in     our    formulation   is  the
condition~(\ref{surfcond}) of vanishing of the surface terms, i.e.~the
fact  that  total derivatives of  fields   or product  of  fields with
respect  to $y$ do  not  contribute to  the  action. Unlike  the naive
Fourier  modes, the mass eigenmode  wavefunctions  $f_n (y)$ and  $g_n
(y)$  comply with   this  crucial  constraint   and hence become   the
appropriate  basis   of       decomposition  of    the   5D     fields
[cf.~(\ref{AmuExpansion})]. Detailed   discussion  is     given     in
Appendix~B, where  we  show  explicitly  that  upon  compactification,
$\Lag_{\rm 5D\GF}$ and $\Lag_{\rm 5D\FP}$ lead  indeed to an effective
4D theory quantized in the conventional $R_\xi$ gauge.

It  is not  difficult to verify  that  the complete 5D QCD  Lagrangian
(\ref{LagYMBKT}) is invariant under the 5D BRS transformations
\begin{eqnarray}
\label{BRS5D}
s A^a_M &=& \ D^{ab}_M\, c^b\;,\nonumber\\
s\, c^a  &=& - \frac{g_5 f^{abc}}{2}\; c^b c^c\;,\\
s\, \overline{c}^a &=& \frac{F[A^a_M]}{\xi}\ .\nonumber
\end{eqnarray}
Following  the    Fourier-like    convolution   method  developed   in
Appendix~C.1, we derive the BRS   transformations for the KK modes  of
the effective theory
\begin{eqnarray}
\label{BRSeff}
s A^a_{(n) \mu} &=& \partial_{\mu} c^a_{(n)}\: -\: g f^{abc} \sum_{m,
l=0}^{\infty} \sqrt{2}^{\; -1-\delta_{n,0}-\delta_{m,0}-\delta_{l,0}}\,
\Delta_{n,m,l} \; A^c_{(m) \mu} c^b_{(l)}\; ,\nonumber\\
s A^a_{(n) 5} &=& - m_n c^a_{(n)}\: -\: g f^{abc} \sum_{m, l=0}^{\infty}
\sqrt{2}^{\; -1-\delta_{l,0}}\, \tilde{\Delta}_{n,m,l} \; A^c_{(m) 5}
c^b_{(l)}\; ,\nonumber\\
s c^a_{(n)} &=& -\; \frac{g f^{abc}}{2} \sum_{m, l=0}^{\infty}
\sqrt{2}^{\; -1-\delta_{n,0}-\delta_{m,0}-\delta_{l,0}}\, \Delta_{n,m,l}
\; c^b_{(m)} c^c_{(l)}\; ,\nonumber\\
s \overline{c}^a_{(n)} &=& \frac{1}{\xi}\, \partial^{\mu} A^a_{(n) \mu}\:
-\: m_n A^a_{(n) 5}\; .
\end{eqnarray}

In close analogy to  our discussion for the  ordinary QCD case in  the
previous  section,  we define  the  generating  functional $Z$ of  the
connected Green functions through the relation:
\begin{eqnarray}
 \label{Z5D}
e^{iZ} &=& \int DA_M \, Dc \, D\overline{c} \ \exp\bigg\{\, i \int d^4x\,
\bigg[\,\int_{-\pi R}^{\pi R} dy\; \Lag_{\YMBKT}\\
&+&\!\!\! \int_{-\pi R}^{\pi R} dy\, [1 + r_c \delta (y)]\, \big(
J^{a M} A^a_M\: +\: \overline{D}^a c^a\: +\:
\overline{c}^a D^a\: 
+ \: K^{a M} s A^a_M\: +\: M^a sc^a\,\big)\, \bigg]\,\bigg\}\; .\nonumber
\end{eqnarray}
The response   of the  generating   functional $Z$  under the  5D  BRS
transformations gives rise to the following master ST identity for the
5D theory:
\begin{equation}
 \label{5DST}
J^{a M} \frac{\delta Z}{\delta K^{a M}}\ -\ \overline{D}^a
\frac{\delta Z}{\delta M^a}\ +\ \frac{1}{\xi}\; D^a\, \partial^\mu
\frac{\delta Z}{\delta J^{a \mu}}\ -\ D^a\,\partial_5\, \frac{\delta
Z}{\delta J^{a 5}} \ =\ 0\, .
\end{equation}
{}From~(\ref{5DST}), we obtain the effective master ST identity
\begin{equation}
\label{masterSTCYM}
\sum_{n=0}^{\infty}\; \bigg( J^{a \mu}_{(n)} \frac{\delta Z}{\delta K^{a
\mu}_{(n)}} + J^{a 5}_{(n)} \frac{\delta Z}{\delta K^{a 5}_{(n)}} -
\overline{D}^a_{(n)} \frac{\delta Z}{\delta M^a_{(n)}} + \frac{1}{\xi}
\partial^{\mu} \frac{\delta Z}{\delta J^{a \mu}_{(n)}} D^a_{(n)} -
m_n\; \frac{\delta Z}{\delta J^{a 5}_{(n)}} D^a_{(n)} \bigg)
\ =\ 0\;.
\end{equation}

As in the ordinary  QCD case, it proves   more practical to  derive ST
identities from the BRS invariance of Green functions. For instance,
we may translate the obvious identity,
\begin{equation}
  \label{green1}
s\, \langle 0 | T \bar{c}^a_{(n)}(x) A^b_{(n) \nu} (z) A^c_{(n) \rho}
(w) A^d_{(n) \sigma} (u) |0 \rangle\ =\ 0\; ,
\end{equation}
into  the corresponding on-shell   ST  identity, which  is graphically
given by
\begin{align}
\begin{split}
\label{onshellST1}
\frac{p_1^{\mu}}{m_n} \hspace{-1mm} \BBBBblob{\mu}{\nu}{\rho}{\sigma}{(n)
p_1}{(n) \; p_2}{(n) \; k_1}{(n) \; k_2} &= \; i \hspace{-7mm}
\SBBBblob{}{\nu}{\rho}{\sigma}{}{}{}{} \hspace{-7mm} +
\frac{p_2^{\nu}}{m_n} \hspace{-5mm} \GGBBblob{}{}{\rho}{\sigma}{}{}{}{}\\
&\phantom{}\\ & \qquad \qquad + \frac{k_2^{\sigma}}{m_n} \hspace{-5mm}
\GBBGblob{}{\nu}{\rho}{}{}{}{}{} \hspace{-7mm} + \frac{k_1^{\rho}}{m_n}
\hspace{-7mm} \GBGBblob{}{\nu}{}{\sigma}{}{}{}{} \hspace{-7mm} \\
&\phantom{}
\end{split}
\end{align}

\vspace{0.5cm}
\noindent
Likewise, the  BRS invariance of  the Green function pertinent  to the
product of  fields $\bar{c}^a_{(n)}(x) A^b_{(n) 5}  (z) A^c_{(n) \rho}
(w) A^d_{(n) \sigma} (u)$, namely
\begin{equation}
  \label{green2}
s\, \langle 0 | T \bar{c}^a_{(n)}(x) A^b_{(n) 5} (z) A^c_{(n) \rho} (w) 
A^d_{(n) \sigma} (u) |0 \rangle\ =\ 0\; ,
\end{equation}
gives rise to following on-shell ST identity:
\begin{align}
\begin{split}
\label{onshellST2}
\frac{p_1^{\mu}}{m_n} \hspace{-1mm} \BSBBblob{\mu}{}{\rho}{\sigma}{(n) \;
p_1}{(n) \; p_2}{(n) \; k_1}{(n) \; k_2} &= \; i \hspace{-7mm}
\SSBBblob{}{}{\rho}{\sigma}{}{}{}{} \hspace{-7mm} -i \hspace{-4mm}
\GGBBblob{}{}{\rho}{\sigma}{}{}{}{}\\ &\phantom{}\\ & \qquad \qquad + \frac{k_2^{\sigma}}{m_n} \hspace{-5mm} \GSBGblob{}{}{\rho}{}{}{}{}{}
\hspace{-7mm} + \frac{k_1^{\rho}}{m_n} \hspace{-7mm}
\GSGBblob{}{}{}{\sigma}{}{}{}{} \hspace{-7mm} \\ &\phantom{}
\end{split}
\end{align}

\vspace{0.5cm}
\noindent
As we will  see in the next section, the  above on-shell ST identities
are important to prove the GET in $2\to 2$ scatterings. In particular,
by  virtue  of  these identities,  we  see  again  that the  KK  modes
$A^a_{(n) 5}$  behave very analogous to the  would-be Goldstone bosons
of     non-linearly     realized     spontaneously    broken     gauge
theories~\cite{Thompson,Ohl}.  A detailed list of Feynman rules, including
would-be Goldstone-boson and ghost  interactions, is given in Appendix
A.


\setcounter{equation}{0}
\section{The 5D Generalized Equivalence Theorem}

In the high-energy  limit of a spontaneously broken  gauge theory, the
amplitude for  emission or  absorption of longitudinal  massive vector
bosons  equals, up to  a phase  and possibly  up to  a renormalization
scheme dependent constant, the amplitude for emission or absorption of
the associated unphysical  would-be Goldstone modes.  This high-energy
equivalence  relation between  the longitudinal  gauge bosons  and the
would-be Goldstone  bosons constitutes the  famous Equivalence Theorem
(ET)~\cite{CLT,LQT}.

In  the  previous  section,  we   have  shown  that  the  on-shell  ST
identities,  which   are  required  for   a  rigorous  proof   of  the
ET~\cite{CG}, are  very analogous to  those of a  spontaneously broken
gauge theory, if the scalar gauge modes $A^{a}_{(n) 5}$ are identified
with the would-be Goldstone bosons.  Such an identification has proven
very important  in order  to establish  the validity of  the ET  in 5D
orbifold Yang--Mills theories without BKTs~\cite{Dicus}. Here, this is
extended to  5D orbifold theories {\em  with} BKTs. In  detail, the ET
reads:
\begin{equation}
  \label{ET}
\begin{split}
T(A^{a_1}_{(n_1) L},\ldots A^{a_k}_{(n_k) L},S \to 
  A^{b_1}_{(m_1) L},\ldots A^{b_l}_{(m_l) L},S')\ & =\ \\ 
C \, i^k \, (-i)^l T(A^{a_1}_{(n_1) 5},\ldots A^{a_k}_{(n_k) 5},S
\to A^{b_1}_{(m_1) 5},\ldots & A^{b_l}_{(m_l) 5},S') 
\ +\  \mathcal{O}(m_{n_i}/E ) \, ,
\end{split}
\end{equation}
where $m_{n_i}$  are the KK masses,  $E$ is the centre  of mass system
(c.m.s.)~energy of the  high-energy scattering process.  The parameter
$C$  is  a constant  that  generally  depends  on the  renormalization
scheme~\cite{Yao}.   However, it  is $C=1$  at the  tree level  and in
certain renormalization  schemes that maintain the  Ward identities of
the   classical  action~\cite{Papa,Denner}.    Finally,  $S,   \,  S'$
collectively denote all particles in the initial and final state which
have no longitudinal  modes.  In order for the  ET~(\ref{ET}) to hold,
the  underlying   gauge  structure  of   the  theory  has   to  ensure
cancellations of all  terms that grow with energy  in the amplitude on
the LHS  of~(\ref{ET}).  As we  will see below,  in higher-dimensional
models with BKTs, such  cancellations are much more intricate, because
the complete infinite tower of the KK modes is involved.

The  ET   is  a  direct   consequence  of   the so-called  generalized
equivalence  theorem (GET)~\cite{CG}.  The GET goes beyond~(\ref{ET}),
by  providing an exact relation  between the relevant amplitudes which
includes  the energetically subleading  terms.   To state  the GET, we
first    note     that   the     longitudinal    polarization   vector
$\epsilon^{\mu}_L$ can be written as
\begin{align}
  \label{longVec}
\epsilon^{\mu}_L(k)\ =\ k^{\mu}/m_n + a^{\mu}(k) \, ,
\end{align}
where $a^{\mu}(k)  = \ord(m_n/E)$ is  the remainder of  the polarization
vector that vanishes in the  high-energy limit.  As an example, we may
consider  the GET  applied to  the production  of two  longitudinal KK
gauge modes:
\begin{eqnarray}
  \label{GETequ}
T(A^a_{(n) T} A^b_{(n) T} \rightarrow A^c_{(n) L} A^d_{(n) L}) &=& \nonumber\\
&&\hspace{-2cm}
-\, T(A^a_{(n) T} A^b_{(n) T} \rightarrow A^c_{(n) 5} A^d_{(n) 5})\:
-\: i T(A^a_{(n) T} A^b_{(n) T} \rightarrow A^c_{(n) 5} a^d_{(n)})\nonumber\\
&&\hspace{-2cm}
-i T(A^a_{(n) T} A^b_{(n) T} \rightarrow a^c_{(n)} A^d_{(n) 5})\: +\:
T(A^a_{(n) T} A^b_{(n) T} \rightarrow a^c_{(n)} a^d_{(n)}) \; .\qquad
\end{eqnarray}
In~(\ref{GETequ}),  $a^a_{(n)}$ indicates    an  amplitude, where  the
longitudinal polarization vector of the  corresponding gauge boson has
been   replaced  by  its  remainder   vector   $a^{\mu}$.  Hence,  the
corresponding amplitudes are  suppressed in the high-energy  limit and
the ET~(\ref{ET}) is recovered.

It is straightforward to apply the  GET to processes with an arbitrary
number   of  longitudinal   gauge     modes.   Specifically,  the  RHS
of~(\ref{GETequ})  may be   found by  performing    the following four
operations:

\begin{itemize}

\item[(i)]  Write down all  amplitudes that  result from replacing any number of longitudinal vector bosons by the respective would-be Goldstone bosons.

\item[(ii)]  Replace  in  each  amplitude the  remaining  longitudinal
polarization vectors by $a^a_{(n)}$.

\item[(iii)]  Multiply  each  of  the resulting  amplitudes  with  the
factors $i^k  (-i)^l$, where $k$  is the number of  the would-be
Goldstone bosons in the initial state and $l$ the corresponding one in
the final state.

\item[(iv)] Sum the complete set of the so-generated amplitudes. 

\end{itemize}
Observe that our GET relation in~(\ref{GETequ}) is consistent with the 
above rules.

Let us present  an explicit proof of  the  GET for the simple  example
of~(\ref{GETequ}), based  on the ST identities   discussed in the last
section.  Using the decomposition, $\epsilon_1 =- k_1/m_n + a_1$, with
$k_1$ being the  incoming momentum, the ST identity~(\ref{onshellST1})
and   the transversality  condition   $\epsilon \cdot   k=0$,  we find
diagrammatically
\begin{align}
  \label{proof2}
&\BBBBblob{\epsilon_a}{\epsilon_b}{\epsilon_1}{\epsilon_2}{(n)p_1}{(n)p_2}
{(n)k_1}{(n)k_2}\quad = \quad -i \hspace{-4mm}
\BBSBblob{\epsilon_a}{\epsilon_b}{}{\epsilon_2}{}{}{}{}\:
+ \:
\BBBBblob{\epsilon_a}{\epsilon_b}{a_1}{\epsilon_2}{}{}{}{}\\
&\phantom{}\nonumber \\ &\phantom{}\nonumber \,
\end{align}
where the  polarization vectors multiplying  the  amplitudes have been
explicitly  displayed at the corresponding  external  leg. 

Our next step consists in making  use of $\epsilon_2=- k_2/m_n + a_2$,
the ST identity~(\ref{onshellST2}), and the fact that $k_1 \cdot a_1 =
m_n$. Substituting all the above  into  the RHS of (\ref{proof2}),  we
obtain
\begin{align}
\begin{split}
\label{proof2general}
\BBBBblob{\epsilon_a}{\epsilon_b}{\epsilon_1}{\epsilon_2}{(n)p_1}
{(n)p_2}{(n)k_1}{(n)k_2}\ = \ - \hspace{-4mm}
&\BBSSblob{\epsilon_a}{\epsilon_b}{}{}{}{}{}{} \hspace{-4mm} +\
\BBGGblob{\epsilon_a}{\epsilon_b}{}{}{}{}{}{}\hspace{-4mm} +\ i\hspace{-4mm}
\BBSBblob{\epsilon_a}{\epsilon_b}{}{a_2}{}{}{}{}\\ &\phantom{}\\
&\phantom{}\\ &\quad -i \hspace{-4mm}
\BBBSblob{\epsilon_a}{\epsilon_b}{a_1}{}{}{}{}{}\hspace{-4mm} -\ 
\BBGGblob{\epsilon_a}{\epsilon_b}{}{}{}{}{}{} \hspace{-4mm} +\
\BBBBblob{\epsilon_a}{\epsilon_b}{a_1}{a_2}{}{}{}{}\\
&\phantom{}\\ &\phantom{}
\end{split}
\end{align}
It    is obvious    that   ghost contributions   cancel,  and    hence
(\ref{proof2general}) is nothing  than (\ref{GETequ}) in  diagrammatic
form.  Following the same reasoning,  one can  prove  the GET for  any
other process.   At   the  tree level,  a proof    based on the   Ward
identities (\ref{wardCYM1})--(\ref{wardCYM5}) is possible but tedious.

Having established  the GET and the resulting  ET, we now show how the
ET  is realized in   a specific tree-level calculation.   Consider the
elastic scattering   of  two longitudinally  polarized  gauge  bosons,
$A^a_{(n)L} A^b_{(n)L} \rightarrow A^c_{(n)L} A^d_{(n)L}$. In terms of
Feynman diagrams, the tree-level amplitude is given by the infinite set
of diagrams
\begin{equation}
\label{gaugeBKT}
\begin{split}
\BBBBblob{\epsilon_a}{\epsilon_b}{\epsilon_c}{\epsilon_d}{(n)p_1}{(n)p_2}
{(n)k_1}{(n)k_2} &= 
\BBBB{\epsilon_a}{\epsilon_b}{\epsilon_c}{\epsilon_d}{(n)p_1}{(n)p_2} 
{(n)k_1}{(n)k_2} +\ \sum_{j=0}^{\infty}
\sChannelB{\epsilon_a}{\epsilon_b}{\epsilon_c}{\epsilon_d}{(n)p_1}{(n)p_2}
{(n)k_1}{(n)k_2}{\,
\, \, \, (j)} \hspace{-1mm} +\ \textrm{crossings}\\ &\phantom{}\\ &=\ 
iT_4+\sum_{j=0}^{\infty}\,
\Big[\,iT^s_{(j)}+iT^t_{(j)}+iT^u_{(j)}\,\Big]\ =\
iT_4+iT^s+iT^t+iT^u \, ,
\end{split}
\end{equation}
where $s$,  $t$  and $u$ refer to  the  pertinent channels. Using  the
Feynman rules of Appendix~A and the relation
\begin{equation}
  \label{epsmu}
\epsilon_a^{\mu}\ =\ \frac{1}{2 m_n \beta} \left[ (1+\beta^2)p_1^{\mu}
- (1-\beta^2) p_2^{\mu} \right] \, ,
\end{equation}
with $\beta = \sqrt{1-4 m_n^2/s}$,  it is straightforward to calculate
the individual diagrams. More explicitly, we find
\begin{eqnarray}
\label{diagramamps}
iT_4 &=& \Delta_{n,n,n,n} \, \frac{i g^2}{8 m_n^4} \bigg[\, f^{abe}
f^{cde} s(t-u)\: +\: f^{ace} f^{bde} t\bigg(s-\frac{u}{\beta^4}\bigg)\: +\:
f^{ade} f^{bce} u\bigg(s-\frac{t}{\beta^4}\bigg)\, \bigg] \, , \nonumber\\
iT^s_{(j)} &=& 2^{-\delta_{j,0}} \Delta^2_{n,n,j} \, i g^2 f^{abe}
f^{cde} \frac{s(u-t)}{8 m_n^4}\, \bigg(\, 1\: +\: \frac{2m_n^2}{s} \bigg)^2 \,
\frac{s}{s-m_j^2} \, , \\ 
iT^t_{(j)} &=& 2^{-\delta_{j,0}}
\Delta^2_{n,n,j} \, i g^2 f^{ace} f^{bde}\, \bigg[\, \frac{u-s}{2t}\, 
\bigg(\, 1\: +\: \frac{t}{2 m_n^2 \beta^2} \bigg)^2\: +\: 
\frac{t-2u}{m_n^2 \beta^2}\, \bigg]\; \frac{t}{t-m_j^2} \, , \nonumber\\ 
iT^u_{(j)} & = & 2^{-\delta_{j,0}}
\Delta^2_{n,n,j} \, i g^2 f^{ade} f^{bce}\, \bigg[\, \frac{t-s}{2u}
\bigg(\, 1\: +\: \frac{u}{2 m_n^2 \beta^2} \bigg)^2 \:
+\: \frac{u-2t}{m_n^2 \beta^2}\, \bigg]\; \frac{u}{u-m_j^2} \, ,\nonumber
\end{eqnarray}
where the Mandelstam variables are given by 
\begin{equation}
s = (p_1+p_2)^2 \, , \quad 
t = (p_1+k_1)^2 =-(1 - c_\theta ) \frac{\beta^2 s}{2} \, , \quad
u = (p_1+k_2)^2 =-(1 + c_\theta) \frac{\beta^2 s}{2} \, , \quad
\end{equation}
and we have used the notation $c_\theta = \cos \theta$.

Our task of verifying the ET  simplifies significantly if the infinite
sums are expanded in terms of $s$. To be precise, using
\begin{equation}
\label{sFactor}
\frac{s}{s-m_j^2}\ =\ 1\: +\: \frac{m_j^2}{s}\: +\: \frac{m_j^4}{s^2}\: +\: 
\frac{m_j^4}{s^2} \frac{m_j^2}{s-m_j^2}\ , 
\end{equation}
the infinite sum $T^{s}$ can be split into four different sums: $T^{s}
= T^{s}_{0} + T^{s}_{2} + T^{s}_{4}  + T^{s}_{6}$, where the subscript
indicates  the power of $m_j$    in the numerator of  (\ref{sFactor}).
Note that  each sum  is convergent.  As  can be  shown by  an explicit
calculation (see end of this  section for a similar case), $T^{s}_{6}$
includes terms  of  order   $m_n/\sqrt{s}$, which  are   energetically
suppressed and do not contribute to the leading part of the amplitude.
Likewise,   we   can make   analogous expansions   for   the $t$-  and
$u$-exchange graphs:
\begin{equation}
\begin{split}
\frac{t}{t-m_j^2}\ &=\ 1 - \left( \frac{2}{1-c_\theta} \, \frac{1}{s} +
\frac{8 m_n^2}{1-c_\theta} \, \frac{1}{s^2} \right) m_j^2 + \left(
\frac{4}{(1-c_\theta)^2} \, \frac{1}{s^2} \right) \, m_j^4\ +\
\ord( m^6_j s^{-3}) \, , \\ 
\frac{u}{u-m_j^2}\ &=\ 1 -
\left(\frac{2}{1+c_\theta} \, \frac{1}{s} + \frac{8m_n^2}{1+c_\theta}
\, \frac{1}{s^2} \right) m_j^2 + \left( \frac{4}{(1+c_\theta)^2} \,
\frac{1}{s^2} \right) \, m_j^4\ +\ \ord(m^6_j s^{-3}) \, .
\end{split}
\end{equation}
The different  infinite sums in $T^{s,t,u}_{0}$,  $T^{s,t,u}_{2}$, and
$T^{s,t,u}_{4}$  can be  calculated  by  means  of (\ref{Snnnn1})  and
(\ref{Snnnn2}), which have been derived  by employing complex analysis
techniques developed in   Appendix~B of~\cite{Paes}. In this   way, we
obtain
\begin{equation}
 \label{T4}
\begin{split}
i T_4\ =\ i g^2\,\Delta_{n,n,n,n} \; \bigg[\, &f^{abe} f^{cde}\, \bigg(
\frac{c_\theta}{8 m_n^4} s^2\ -\ \frac{c_\theta}{2 m_n^2} s\, \bigg)\\
&+\, f^{ace} f^{bde}\, \bigg(\,
\frac{(c_\theta+3)(c_\theta-1)}{32m_n^4} s^2\ +\
\frac{1-c_\theta}{4m_n^2} s\, \bigg)\\ &+\, f^{ade} f^{bce}
\bigg(\,\frac{(c_\theta-3)(c_\theta+1)}{32m_n^4} s^2\ +\
\frac{c_\theta+1}{4m_n^2} s \bigg)\, \bigg]\; ,
\end{split}
\end{equation}
\begin{eqnarray}
  \label{Ts}
iT^s &=& ig^2 \; f^{abe} f^{cde} \; \bigg[\, \Delta_{n,n,n,n}\,
\bigg( -\, \frac{c_{\theta}}{8 m_n^4} s^2\: -\: 
\frac{c_{\theta}}{6 m_n^2} s\: +\:
\frac{5 c_{\theta}}{4}\, \bigg)\nonumber\\
 &&\hspace{3cm} -\ \frac{c_{\theta}}{2} X_n\, \bigg]\ +\
\ord(m_n/\sqrt{s})\; ,\\[4mm]
  \label{Tt}
iT^t &=& ig^2 \; f^{ace} f^{dbe} \; \bigg[\, \Delta_{n,n,n,n}\, \bigg(\,
\frac{(c_{\theta}+3)(1-c_{\theta})}{32 m_n^4} s^2\: +\: \frac{11
c_{\theta}-3}{12 m_n^2} s\: +\: \frac{8
c_{\theta}^2-5c_{\theta}+9}{6(1-c_{\theta})}\, \bigg)\nonumber\\ 
&&\hspace{3cm} +\; \frac{3+c_{\theta}}{2(1-c_{\theta})}\, X_n\bigg]\ 
+\ \ord(m_n/\sqrt{s})\; ,\\[4mm]
  \label{Tu}
iT^u &=& ig^2 \; f^{ade} f^{bce} \; \bigg[\, \Delta_{n,n,n,n}\, \bigg(\,
\frac{(3-c_{\theta})(1+c_{\theta})}{32 m_n^4} s^2\: -\: \frac{11
c_{\theta}+3}{12 m_n^2} s\: 
+\: \frac{8 c_{\theta}^2+5c_{\theta}+9}{6(1+c_{\theta})}\, \bigg)\nonumber\\ 
&&\hspace{3cm} +\; \frac{3-c_{\theta}}{2(1+c_{\theta})}\, X_n\, \bigg]\ 
+\ \ord(m_n/\sqrt{s})\, ,
\end{eqnarray}
where
\begin{equation}
  \label{Xn}
X_n\ =\ 8\, N_n^4\,\pi^2 R^2 \tilde{r}_c^3 m_n^2 \; .
\end{equation}
Collecting all contributions  (\ref{T4})--(\ref{Tu}), we find that the
$s^2$-contributions in  the $s$-,  $t$- and $u$-channel  graphs cancel
against terms  in $T_4$.  Terms linear  in $s$ are  identical for each
colour factor.   Thus, they  vanish due to  the Jacobi  identity.  The
final result is then given by
\begin{align}
\label{LOgaugeBKT}
iT_4&+iT^s+iT^t+iT^u\\
&=\ i g^2\,(\Delta_{n,n,n,n} + X_n)\,  \bigg[\, f^{ace} f^{dbe}
\frac{c^2_\theta + 3}{2 (c_\theta - 1)}\: 
+\: f^{ade} f^{bce} \frac{c^2_\theta + 3}{2(c_\theta + 1)}\, \bigg]\ +\
\ord ( m_n/\sqrt{s})\; . \nonumber
\end{align}
In  (\ref{LOgaugeBKT}),  we  only  exhibit  the  leading  contribution
$\ord(1)$, which  will be essential  for checking the validity  of the
ET.

Let us  now   consider the  RHS  of  the  ET~(\ref{ET}).  The   scalar
scattering amplitude is given by
\begin{equation}
\begin{split}
  \label{scalarBKT}
\SSSSblob{}{}{}{}{(n)p_1}{(n)p_2}{(n)k_1}{(n)k_2} \hspace{-2mm}\ &=\
\sum_{j=0}^{\infty}
\sChannelS{}{}{}{}{(n)p_1}{(n)p_2}{(n)k_1}{(n)k_2}{\, \, \, \, (j)}
\hspace{-2mm}\ +\quad \textrm{crossings}\\ &\phantom{}\\ & =\
\sum_{j=0}^{\infty}\, \Big[\, iT^s_{(j)}+iT^t_{(j)}+iT^u_{(j)}\,
\Big]\ =\ iT^s+iT^t+iT^u\, .
\end{split}
\end{equation}
Using  the Feynman rules of  Appendix~A,  we obtain for the individual
$s$-, $t$- and $u$-channel graphs
\begin{equation}
\begin{split}
  \label{Scalar}
iT^s_{(j)}\ &=\ i g^2\: 2^{-\delta_{j,0}} \tilde{\Delta}^2_{n,j,n} \, 
f^{abe} f^{cde}\; \frac{u-t}{2(s-m_j^2)} \ , \\ 
iT^t_{(j)}\ &=\ i g^2\: 2^{-\delta_{j,0}} \tilde{\Delta}^2_{n,j,n} \, 
f^{ace} f^{dbe}\; \frac{s-u}{2(t-m_j^2)} \ , \\ 
iT^u_{(j)}\ &=\ i g^2\:
2^{-\delta_{j,0}} \tilde{\Delta}^2_{n,j,n} \, f^{ade} f^{bce}\;
\frac{t-s}{2(u-m_j^2)} \ .
\end{split}
\end{equation}
Considering      the      analytic   form       of      the  couplings
$\tilde{\Delta}_{n,j,n}$ in (\ref{Scalar}),  it turns out that only  a
single expansion in terms of $1/s$, i.e.
\begin{equation}
  \label{sumsj}
\frac{s}{s-m_j^2}\ =\ 1\: +\: \frac{m_j^2}{s-m_j^2} \, ,
\end{equation}
will  be sufficient  to evaluate  the leading  part of  the amplitude.
Using   (\ref{Snnnn2}),  we   can  now   sum  the   leading   part  of
(\ref{Scalar}).    Up  to   energetically   subleading  terms   $\ord(
m_n/\sqrt{s})$,     the     amplitude     (\ref{scalarBKT})     equals
(\ref{LOgaugeBKT}).  This completes  our check  of the  ET (\ref{ET}).
For vanishing  $r_c$, i.e.~$r_c \to 0$, we  find $\Delta_{n,n,n,n} \to
3$ and $X_n \to 0$, and  so recover the result in~\cite{Dicus} for the
validity of the ET in 5D orbifold theories without BKTs.

We conclude  this section by showing  that  the infinite sum involving
the second  term   in~(\ref{sumsj}) is   indeed  subleading  of  order
$m_n/\sqrt{s}$.  We  modify  the second equation  of (\ref{Snnnn2}) by
taking  the  additional factor $m_j^2/(s-m_j^2)$   into account.  With
this modification, an explicit calculation gives
\begin{equation}
\label{surfaceSum}
\begin{split}
\sum_{j=0}^{\infty} 2^{-\delta_{j,0}}\, \tilde{\Delta}^2_{n,j,n}\;
\frac{m_j^2}{s-m_j^2}\ =\ \, &\tilde{\Delta}_{n,n,n,n}\, \frac{4
m_n^2}{s-4 m_n^2}\\ &\hspace{-2cm}+\; 32 \pi^2 R^2 m_n^2 \tilde{r}_c^3
N_n^4 \, \Big(\, 1 - \pi^2 R^2 m_n^2 \tilde{r}_c^2\, \Big)\, \bigg[\,
\frac{m_n^2}{s-4 m_n^2} + \frac{m_n^4}{(s-4 m_n^2)^2}\,\bigg]\\
&\hspace{-2cm}+\, \frac{8 \pi^4 R^4 m_n^4 \tilde{r}_c^5 N_n^4}{s+
\frac{\sqrt{s}}{\pi R \tilde{r}_c} \tan \pi R \sqrt{s}} \,
\frac{s^3-4m_n^2 s^2 + 4 m_n^4 s}{(s-4 m_n^2)^2}\ -\ 8 \pi^4 R^4 m_n^4
\tilde{r}_c^5 N_n^4 \; .
\end{split}
\end{equation}
The first two terms are obviously $\ord(m^2_n/s)$.  The third term has
got poles at the spectrum and we  therefore take the high-energy limit
in a discrete manner:  $\sqrt{s_n} =  (n-\frac{1}{4})/R$, with $n  \to
\infty$.   The third term approaches the  negative of the fourth term.
Consequently, all $\ord(1)$  terms cancel and (\ref{surfaceSum})  does
contribute only subleading terms of order $m_n/\sqrt{s}$.


\setcounter{equation}{0}
\section{High Energy Unitarity Bounds}
\label{bounds}
\setcounter{equation}{0}

In   higher-dimensional  field theories,  the  coupling  constants are
dimensionful parameters. For example, for the  case of 5D $N_c$-colour
QCD, the coupling $g_5$   has the energy  dimensions   $E^{-1/2}$.  On
grounds of naive dimensional analysis, the $s$-wave amplitude $a_0$ of
a $2 \to 2$ scattering involving 5D gluons behaves as
\begin{equation}
  \label{a0E5}
a_0 (E_5) \ \sim \ N_c\, g^2_5\, E_5\; ,
\end{equation}
where $E_5$ is the  c.m.s.~energy  in 5D.   This means that  at energy
scales  $E_5   \stackrel{>}{{}_\sim} 1/(g^2_5\,N_c)$, the transition
amplitude $a_0$  exceeds 1, thereby invalidating  perturbation theory.
In  addition,  the fact that  the  asymptotic high-energy behaviour of
$a_0$ does not  approach  a constant,  but   grows linearly with   the
energy,      is  in gross    violation     with  the Froissart--Martin
bound~\cite{Froissart}.

In  a  theory compactified to  4D,  the 3- and   4-gluon couplings are
dimensionless, while the  KK spectrum is infinite.  In this  case, the
aforementioned   energy upper  limit   due  to unitarity  violation is
translated into a  corresponding upper limit on the  number  of the KK
modes.  To     be  specific, one  gets    the   very approximate upper
bound~\cite{Dicus}
\begin{equation}
\frac{N_0}{R} \ \stackrel{<}{{}_\sim} \ \frac{1}{g^2_5\,N_c }\ .
\end{equation}
Obviously,   this   upper   bound   depends   non-trivially   on   the
compactification radius $R$, the 5D gauge-coupling constant $g_5$ and
the number of colours $N_c$.

We will now compute  explicitly  the perturbative unitarity limits  on
our 5D  orbifold theory with one BKT.   For this purpose, let us first
describe our  approach to perturbative  unitarity.  As usual, we start
by stating the optical theorem for a general $i \to f$ transition
\begin{equation}
  \label{OT}
2\, \operatorname{Im} T_{f i}\ =\ \sum_j \; T_{f j}\, T^*_{j i}\; ,
\end{equation}
where the sum over $j$ is understood to include phase-space correction
factors  for   each  of  the   individual  states  in   $j$.
Decomposing  the transition  amplitude $T_{fi}$  in terms  of Legendre
polynomials~\cite{Barton},
\begin{equation}
  \label{pwExp}
T_{fi}(s,c_\theta)\ =\ 16 \pi\, \sum_{l=0}^{\infty}\, (2l+1)\, 
P_l(c_\theta)\; a_l (s)\; ,
\end{equation}
with  $a_l   (s)  =  1/(32\pi)\,   \int_{-1}^{+1}\,  dc_\theta\,
T_{fi} P_l(c_{\theta})$,  and   substituting  the  resulting  expression~(\ref{pwExp})
into~(\ref{OT}),  we  obtain a  unitarity  relation  for the  $s$-wave
($l=0$) amplitudes
\begin{equation}
  \label{a0OT}
{\rm Im}\, [a_0]_{f i}\ =\  \sum_j \; \sigma_j \; [a_0]_{f j} \;
[a_0]^*_{j i} \; .
\end{equation}
Here,  we  denote  with  $\sigma_j =  \lambda(s,  m_1^2,m_2^2)/s$  the
phase-space  factors for  2-particle states,  with $\lambda  (x,y,z) =
[ x^2 + y^2 + z^2 - 2(x y + y z + z x) ]^{1/2}$.

Our approach    to perturbative unitarity  consists  in  absorbing the
phase-space factors $\sigma_j$ into  the definition of  the transition
amplitudes $[a_0]_{i j}$, i.e.
\begin{equation}
 \label{a0tilde} 
[\tilde{a}_0]_{ij}\ =\ \sqrt{\sigma_i \sigma_j} \; [a_0]_{ij}\; .
\end{equation}
This enables us to cast~(\ref{a0OT}) in the simple matrix form
\begin{equation}
  \label{OTtilde}
{\rm Im}\, \tilde{a}_0\ =\ \tilde{a}_0\, \tilde{a}_0^*\; .
\end{equation}
Since our unitarity limits will be deduced  from $2 \to 2$ scatterings
that involve longitudinal KK gauge modes,  $\tilde{a}_0$ will turn out
to  be   a square symmetric   matrix.  Therefore, $\tilde{a}_0$ can be
diagonalized by means of a {\em real} orthogonal matrix $R$ as
\begin{equation}
  \label{ROT}
R^T\, \tilde{a}_0\, R \ =\ \hat{a}_0\;, 
\end{equation}
where   $\hat{a}_0$ is a diagonal   transition  amplitude matrix.  The
unitarity   relation~(\ref{OTtilde})  implies   that not          only
$\tilde{a}_0$, but also  its  imaginary part ${\rm Im}\,  \tilde{a}_0$
can  be diagonalized by the same  orthogonal matrix $R$. Consequently,
the following unitarity relation may be established:
\begin{equation}
  \label{OTrelation}
{\rm Im}\, \hat{a}_0\ =\ \big({\rm Re}\,\hat{a}_0\big)^2\: 
  +\: \big({\rm Im}\,\hat{a}_0\big)^2\;.
\end{equation}
Equation~(\ref{OTrelation}) directly implies the inequality
\begin{equation}
  \label{PUL}
{\rm Im}\, \hat{a}_0\ \leq \ {\bf 1}\; ,
\end{equation}
which is  equivalent, by means of~(\ref{OTtilde}),  to the requirement
that  the  largest  eigenvalue   of  $\hat{a}_0$  or  equivalently  of
$\tilde{a}_0$ (in absolute value terms) should smaller than 1.  Notice
that  our approach  to  perturbative unitarity  differs  from the  one
applied  originally  for the  SM  case~\cite{LQT},  in  the fact  that
phase-space effects were not  considered in the latter.  However, such
effects  are  non-negligible for  the  heavier  KK  modes, and  should
therefore be included consistently.

Since  the 5D  transition  amplitudes diverge  with increasing  energy
[cf.~(\ref{a0E5})],   the    unitarity   constraint~(\ref{PUL})   will
explicitly depend  on the c.m.s.~energy $\sqrt{s}$.   For this reason,
perturbative  unitarity limits  will  constrain not  only the  maximum
allowed  number  of  KK   modes,  e.g.~$N_{\rm  max}$,  but  also  the
c.m.s.~energy, i.e.~$\sqrt{s}_{\rm  max} = 4m^2_{N_{\rm  max}}$, above
which unitarity  is violated.   To avoid complications  arising mainly
from   IR   singularities   in    the   KK   gluon   scatterings,   we
follow~\cite{Dicus}  and  perform a  coupled  channel analysis  (CCA),
based on the inelastic colour-singlet processes $A^a_{(n)5} A^a_{(n)5}
\to A^b_{(m)5} A^b_{(m)5}$, with $n\neq  m$ and $n,m \leq N_{\rm max}$
and $s =  4 m^2_{N_{\rm max}}$.  Strictly speaking,  our CCA relies on
the   physical  processes   $A^a_{(n)L}   A^a_{(n)L}  \to   A^b_{(m)L}
A^b_{(m)L}$,  but  we  have  made   use  of  the  ET  to  replace  the
longitudinal  KK  gauge  bosons  $A^a_{(n)L}$  with  their  respective
would-be  Goldstone bosons  $A^a_{(n)5}$.  Such  a consideration  is a
good approximation for the lighter  KK modes, with masses much smaller
than  the c.m.s~energy  $\sqrt{s}$.  For  the heavier  KK  modes, with
masses $m_{(n)} \sim \sqrt{s}$, our simplification may be justified by
the fact  that scatterings  of those heavy  states will  be relatively
suppressed  by  phase-space  factors  that  occur  in  the  transition
amplitudes                 $[\tilde{a}_0]_{nm}$                defined
in~(\ref{a0tilde}). Nevertheless,  our derived upper  bounds should be
regarded to  be slightly conservative, in  the sense that  we have not
taken into account $2\to 2$  scatterings where all asymptotic KK gluon
states have different masses.  However, initial estimates convinced us
that     these    contributions    to     the    CCA     are    really
negligible.\footnote{These estimates  will be further  consolidated by
our  analysis of  unitarity bounds  from the  decay widths  of  the KK
states, which is given at the end of the section.}

The phase-space-corrected $s$-wave amplitudes $[\tilde{a}_0]_{nm}$ for
the $2\to 2$   inelastic   scatterings, $A^a_{(n)5} A^a_{(n)5}     \to
A^b_{(m)5} A^b_{(m)5}$, are given by
\begin{equation}
  \label{a0jSum}
[\tilde{a}_0]_{nm}\ =\ \frac{S_{n,m}}{s}  
\sum_{j=0}^{\infty}\; [a^{(j)}_0]_{nm}\; ,
\end{equation}
where   $S_{n,m}  =    \sqrt{(s-4m_n^2)(s-4m_m^2)}$  are   phase-space
correction factors  and $[a^{(j)}_0]_{nm}$ are the individual $s$-wave
amplitudes mediated by the exchange of the $j$ KK gauge mode $A^c_{(j)
\mu}$:
\begin{eqnarray}
[a_0^{(j)}]_{nm} &=& -\, \frac{g^2 N_c}{32 \pi}\; \tilde{\Delta}_{n,j,m}^2 \;
2^{-\delta_{j,0}}\nonumber\\ 
&&\times\; \bigg[\, 1\: +\: \frac{2 (s - m_n^2 - m_m^2) + m_j^2}{S_{n,m}}\; 
\ln\bigg| \frac{s + 2(m_j^2 - m_n^2 - m_m^2) - S_{n,m}}{s + 
2 (m_j^2 - m_n^2 - m_m^2) + S_{n,m}} \bigg|\; \bigg]\; .\qquad
\end{eqnarray}
It is interesting to observe that in the high-energy  limit $s \gg (m_n
+   m_m)^2$,   the   $s$-wave amplitudes  $[a_0^{(j)}]_{nm}$   diverge
logarithmically,
\begin{equation}
[a^{(j)}_0]_{nm} \ = \ -\, \frac{g^2 N_c}{32 \pi}\,
\tilde{\Delta}_{n,j,m}^2 \; 2^{-\delta_{j,0}}\; \bigg[\, 1\: -\: 2 \ln
\bigg(\frac{s}{m_j^2}\bigg)\, \bigg]\ +\ {\cal O}\bigg(\frac{(m_n +
m_m)^2}{s}\bigg) \; .
\end{equation}
In the limit $r_c \to 0$, the terms $j=n+m$ and $j=|n-m|$ dominate the
sum    (\ref{a0jSum}) and  the result~\cite{Dicus}    of a 5D
orbifold theory without BKTs is recovered
\begin{equation} 
[a_0]_{nm}\ =\ -\,
\frac{g^2 N_c}{32 \pi}\, \bigg[\, 1\: -\: 
2 \ln \bigg(\frac{s}{|n^2/R^2-m^2/R^2|}\bigg)\, \bigg]\; .
\end{equation}
Because  of the  presence of  the BKT,  our CCA involves  the infinite
sum~(\ref{a0jSum}), thus making  it technically  more challenging than
the corresponding analysis without BKTs.  We perform this infinite sum
and  find the maximum  eigenvalue $\lambda_{\rm max}$ of $\tilde{a}_0$
numerically.

\begin{figure}[t]
\begin{center}
\includegraphics[width=0.48\textwidth]{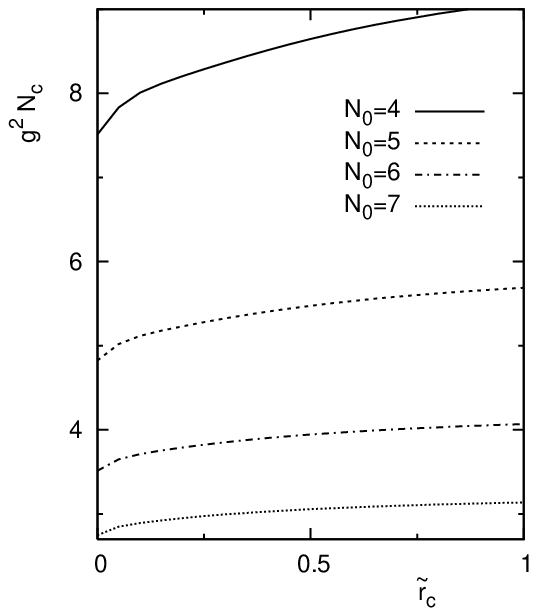}
\includegraphics[width=0.48\textwidth,]{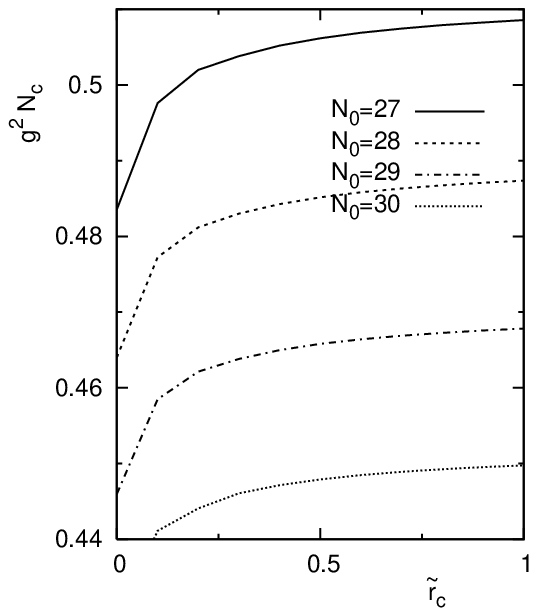}
\caption{\em Contour plots in the  plane $(\tilde{r}_c = r_c/(2\pi R),
g^2N_c)$, for  fixed values of   $N_{\rm max}$.   The  area above  the
contour lines is excluded by perturbative unitarity.}\label{contour}
\end{center}
\end{figure}

In  Figure~\ref{contour},  we  display  contour  plots  in  the  plane
$(\tilde{r}_c, g^2  N_c)$, for different  fixed values of  the maximum
KK-mode  number $N_{\rm  max}$,  which is  obtained  by requiring  the
perturbative unitarity constraint:  $|\lambda_{\rm max}| \leq 1$.  The
area that  lies above  the contour lines  is excluded  by perturbative
unitarity.  For the  case without BKTs, $\tilde{r}_c =  r_c/(2\pi R) =
0$, our  upper limits  are larger by  a about  of factor 2  than those
presented  in~\cite{Dicus}.   The origin  of  this  difference may  be
traced to the effect of the phase-space factors $\sigma_j$, which have
not been included in the analysis of~\cite{Dicus}.  In the presence of
BKTs, we find that the unitarity bounds show only a weak dependence on
the  size  $\tilde{r}_c$ of  the  BKTs.   In  particular, the  maximum
allowed value of $g^2  N_c$ thanks to perturbative unitarity increases
only by  about $5\%$.   This behaviour may  be understood by  the fact
that  high-energy unitarity  probes  distances much  smaller than  the
compactification radius $R$ where the  effect of the BKTs becomes less
significant.

\begin{figure}[t]
\begin{center}
\includegraphics[width=0.70\textwidth]{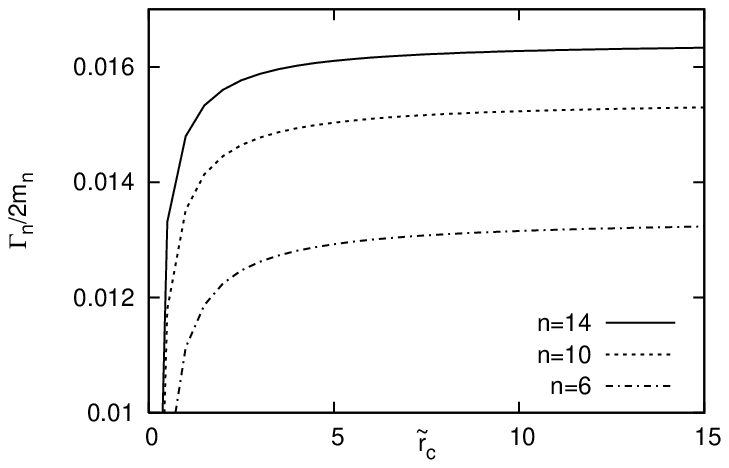}\\ \vspace{3mm}
\includegraphics[width=0.70\textwidth]{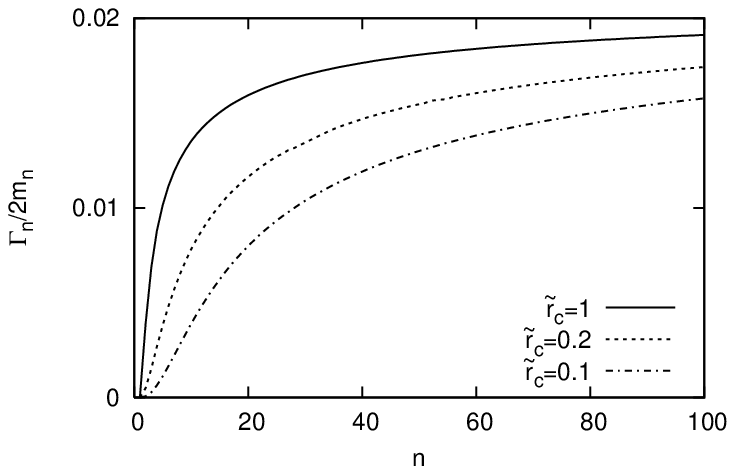}
\caption{\em Numerical values for $\Gamma_n/2 m_n$ as functions of 
$\tilde{r}_c$ and $n$}\label{decayWidths}
\end{center}
\end{figure}

An interesting consequence  of the BKT is that the  KK gauge modes are
no longer stable. One would  have naively expected that the larger the
size $\tilde{r}_c$ of the BKT is, the larger the decay width of the $n$th KK
gauge  boson becomes,  thereby  leading to  a  potential violation  of
perturbative unitarity.  However, we will  show that this is {\em not}
the case, in agreement with our results from the $2\to 2$ scatterings.
Perturbative unitarity in the decays of the KK gauge bosons gives rise
to the constraint,
\begin{equation}
\label{widthCond}
\frac{1}{2}\,\Gamma_n\  \leq\ m_n\ ,
\end{equation}
where $\Gamma_n$ is the total decay width of the  $n$th KK gauge boson
$A^a_{(n)\mu}$.    At the lowest order    of perturbation theory, only
two-body  decays of $A^a_{(n)\mu}$ into  lighter  KK gauge bosons will
contribute. The  total decay width  $\Gamma_n$ is then  the sum of the
partial widths and may conveniently be expressed as
\begin{equation}
  \label{widthsSum}
\Gamma_n\ =\ \frac{1}{16 \pi m_n^3}\ \sum_{k,l=0 \; (k+l \leq n)} 
\lambda (m^2_n, m^2_k, m^2_l)\ |T_{(n,k,l)}|^2\ ,
\end{equation}
where    $T_{(n,k,l)}$    is  the   transition   amplitude   for decay
$A^a_{(n)\mu}  \to  A^b_{(k)\mu}   A^c_{(l)\mu}$, with    $n  \ge    k
+l$. Squaring $T_{(n,k,l)}$ and averaging over initial states, we find
\begin{eqnarray}
  \label{pwidth}
|T_{(n,k,l)}|^2 & =& \frac{1}{3} g^2 N_c \pi^2 R^2 \tilde{r}_c^3 N_n N_k
N_l\; \Delta_{n,k,l}\nonumber\\
&&\times\, \Big[\, m_n^4 + m_k^4 + m_l^4 +
10\big(\, m_n^2 m_k^2 + m_k^2 m_l^2 +m_l^2 m_n^2\,\big)\, \Big]\, .
\end{eqnarray}
Here, we should note that not all kinematically allowed decay channels
of  $A^a_{(n)\mu}$  contribute    to  $\Gamma_n$.   For   example, the
transition amplitudes $|T_{(n,k,0)}|$, with  $0  \leq k < n$,   vanish
identically.

In Fig.~\ref{decayWidths},  we  present   numerical estimates  of  the
quantity $\Gamma_n/(2m_n)$,   as   functions   of  the   BKT  coupling
$\tilde{r}_c$ and the  KK  number $n$  of the decaying
gauge  boson $A^a_{(n)\mu}$.  After its rapid   increase, the value of
$\Gamma_n/(2m_n)$  saturates to  values much smaller  than  1 at large
$\tilde{r}_c$  and/or large $n$. Consequently, we  find as before that
BKTs do not   lead  by  themselves  to  a  violation  of  perturbative
unitarity in the decays of the KK gauge bosons.


\setcounter{equation}{0}
\section{Conclusions}
\label{conclusion}
\setcounter{equation}{0}

We have studied the quantization and high-energy unitarity of 5D field
theories compactified on  an $S^1/\mathbb{Z}_2$ orbifold  that include
localized gauge-kinetic  terms at  the  orbifold fixed points.   These
localized interactions, the  so-called BKTs, emerge naturally in these
theories as counterterms at tree-level to absorb  the UV infinities that
are generated at the quantum level from operators of the same form.

Extending our approach to quantization in~\cite{MPR}, we have
developed a  functional  differentiation   formalism to  quantize  the
theory within the framework of generalized  $R_\xi$ gauges, before the
KK reduction.  With the aid of this formalism, we  were able to derive
the  master  Ward  and  ST identities  which  is a  reflection of  the
underlying gauge  and BRS symmetries  of the theory.   The Ward and ST
identities  relate the divergence of  a  given Green function to other
Green functions of the theory.  By virtue of  such identities, we have
stated a generalized form of the ET, which we call the 5D GET.  The 5D
GET extends the equivalence relation between the longitudinal KK gauge
modes  and their respective would-be  Goldstone bosons to consistently
include the energetically suppressed terms in high-energy scatterings.

An important  property of  any  perturbative predictive  framework  of
field theory    is  perturbative  unitarity.    Requiring perturbative
unitarity, we have deduced upper limits on  the number of the KK modes
as functions of the size  of the gauge  coupling and the colour of the
5D Yang--Mills theory.  Our approach  to perturbative unitarity  takes
into consideration phase space corrections due to heavier KK modes and
so   improves upon  earlier studies  on   the same topic~\cite{Dicus}.
Because of this, the derived upper limits have been found to be weaker
by a factor  $\sim 2$ than those obtained  in these  studies.  We have
investigated the impact  of the BKTs on   these unitarity limits.   We
have shown that the unitarity limits weakly  depend on the size of the
BKTs.  Such a  behaviour may be attributed to   the fact that  at high
energies   distances  smaller than    the compactification  radius are
probed.  As a consequence,  high-energy unitarity strongly depends  on
the  bulk dynamics of the  theory, and therefore high-energy unitarity
limits show a weak dependence on the size of the BKTs.

Even though large  BKTs may weakly affect the  unitarity limits on the
theory, they can, however, make  the higher KK modes decouple from the
dynamics of the lowest lying KK states~\cite{Carena}, e.g.~that of the
gluons.  Their presence may  help to considerably alleviate the severe
phenomenological constraints that apply  to those theories from a full
fledged analysis of the electroweak data~\cite{MPR2}.  It is therefore
of great phenomenological interest  to analyze in detail the prospects
of  probing such  BKT-dominated  theories  at the  LHC  and at  future
$e^+e^-$ linear colliders.

\subsection*{Acknowledgements}    
We wish  to thank  Alex Pomarol  for discussions, as well as
Francisco del Aguila, Jose Santiago and Manuel Perez-Victoria for
pointing out Ref.~\cite{Santiago} to us.  The  work of  AP is
supported in part by  the PPARC research grant PPA/G/O/2002/00471. The
work of AM and RR  is supported by the Bundesministerium f\"ur Bildung
und  Forschung  (BMBF,  Bonn,   Germany)  under  the  contract  number
05HT4WWA2.


\newpage
\def\theequation{\Alph{section}.\arabic{equation}}
\begin{appendix}

\setcounter{equation}{0}
\section{Feynman Rules}\label{rules}

The two-point functions for the  KK mass eigenstates, deduced from the
classical part of the 5D Lagrangian~(\ref{LagYMBKT}), i.e.~without the
terms $\Lag_{\rm 5D\GF}$ and $\Lag_{\rm 5D\FP}$, are given by:
\begin{align*}
\BB{a \; \mu}{b \; \nu}{(n) \; p}
&\parbox{130mm}{
\begin{equation}
\label{twopoint}
\Gamma^{a b}_{\mu \nu} (p_{(n)})\ =\ i \delta^{ab} \big[ 
-g_{\mu \nu}\, \big( p^2 - m_n^2\big)\: +\: 
p_{\mu} p_{\nu}\, \big]\; ,
\end{equation}}\\
\BS{a \; \mu}{b}{(n) \; p}
&\parbox{130mm}{
\begin{equation}
\label{mixing2}
\Gamma^{a b}_{\mu 5} (p_{(n)})\ =\ \delta^{ab} m_n p_{\mu}\; ,
\end{equation}}\\
\SSLL{a}{b}{(n) \; p}
&\parbox{130mm}{
\begin{equation}
\Gamma^{a b}_{5  5} (p_{(n)})\ =\ i \delta^{ab} p^2\; .
\end{equation}}\\
\end{align*}
The  two-point function~(\ref{mixing2})  refers  to the  non-vanishing
mixing between vector  and scalar modes that occurs  in the absence of
the    gauge-fixing    term    $\Lag_{\rm   5D\GF}$    discussed    in
Section~\ref{wardCYMBKT}.   In  the  presence of  $\Lag_{\rm  5D\GF}$,
however, the two-point functions take on the form:
\begin{align*}
\BB{a \; \mu}{b \; \nu}{(n) \; p}
&\parbox{130mm}{
\begin{equation}
\label{twopointGF}
\Gamma^{ab}_{\mu \nu} (p_{(n)},q_{(n)})\ =\ i \delta^{ab} \big[ 
-g_{\mu \nu}\, ( p^2 - m_n^2)\: +\: 
\big( 1 - \mbox{$\frac{1}{\xi}$}\big)\, p_{\mu} p_{\nu}\,\big]\; ,
\end{equation}}\\
\SSLL{a \; \mu}{b}{(n) \; p}
&\parbox{130mm}{
\begin{equation}
\Gamma^{ab}_{55} (p_{(n)},q_{(n)})\ =\ i \delta^{ab} \big( p^2\: -\: \xi
m_n^2 \big)\; ,
\end{equation}}\\
\GG{a}{b}{(n) \; p}
&\parbox{130mm}{
\begin{equation}
\Gamma^{ab}_{\bar{c} c} (p_{(n)},q_{(n)})\ =\ -i \delta^{ab} \big( p^2\: -\:
\xi m_n^2\, \big)\; .
\end{equation}}
\end{align*}
Correspondingly, the  propagators for the vector, scalar  and ghost KK
modes read:
\begin{align}
\label{propagator}
D^{ab}_{\mu \nu} (p_{(n)})\ &=\ \frac{i
\delta^{ab}}{p^2-m_n^2+i\epsilon}\ \bigg[ -\,g_{\mu \nu}\: +\: (1- \xi)\,
\frac{p_{\mu} p_{\nu}}{p^2-\xi m_n^2}\, \bigg]\ ,\\ 
D^{ab}_{55} (p_{(n)})\ &=\
\frac{i \delta^{ab}}{p^2- \xi m_n^2 +i \epsilon}\ ,\\
D^{ab}_{\overline{c} c} (p_{(n)})\ &=\ \frac{-i \delta^{ab}}{p^2- \xi
m_n^2 +i \epsilon}\ .
\end{align}
Using the definitions  introduced in Appendix~\ref{multiplication} for
the    effective    gauge-coupling   coefficients    $\Delta_{n,m,l}$,
$\tilde{\Delta}_{n,m,l}$ etc., the  Feynman rules for the interactions
of the KK mass eigenstates are given by
\begin{align*}
\BBBeast{a \; \mu}{b \; \nu}{c \; \rho}{(n) \; k}{(m) \; p}{(l) \; q}
&\parbox{130mm}{
\begin{equation}
\begin{split}
\label{triple}
\Gamma^{abc}_{\mu \nu \rho} &(k_{(n)},p_{(m)},q_{(l)})\ =\ \\ &g f^{abc}
\sqrt{2}^{\; -1-\delta_{n,0}-\delta_{m,0}-\delta_{l,0}} \,
\Delta_{n,m,l}\\ &\times \big[g_{\mu \nu}(k-p)_{\rho}+g_{\rho
\mu}(q-k)_{\nu}+g_{\nu \rho}(p-q)_{\mu} \big]\; ,
\end{split}
\end{equation}}\\ &\phantom{}\\
\BBSeast{b \; \mu}{c \; \nu}{a}{(m) \; p}{(l) \; q}{(n) \; k}
&\parbox{130mm}{
\begin{equation}
\begin{split}
\Gamma^{abc}_{5 \mu \nu} &(k_{(n)},p_{(m)},q_{(l)})\ =\ i g f^{abc} \,
g_{\mu \nu}\\ & \times \big[ m_l \sqrt{2}^{\; -1-\delta_{m,0}}
\tilde{\Delta}_{n,m,l} - m_m \sqrt{2}^{\; -1-\delta_{l,0}}
\tilde{\Delta}_{n,l,m} \big]\; ,
\end{split}
\end{equation}}\\ &\phantom{}\\
\SSBeast{b}{c}{a \; \mu}{(m) \; p}{(l) \; q}{(n) \; k}
&\parbox{130mm}{
\begin{equation}
\Gamma^{abc}_{\mu 5 5} (k_{(n)},p_{(m)},q_{(l)})\ =\ g f^{abc}
\sqrt{2}^{\; -1-\delta_{n,0}} \tilde{\Delta}_{l,n,m} (q-p)_{\mu}\; ,
\end{equation}}\\
\end{align*}
\begin{align*}
\BBBB{a \; \mu}{b \; \nu}{c \; \rho}{d \; \sigma}{(n) \; k}{(m) \;
p}{(l) \; q}{(k) \; r} &\parbox{130mm}{
\begin{equation}
\begin{split}
\Gamma^{abcd}_{\mu \nu \rho \sigma} (&k_{(n)},p_{(m)},q_{(l)},r_{(k)})
\ =\ \\ i g^2 &\Delta_{n, m, l, k} \sqrt{2}^{\;
 -2-\delta_{n,0}-\delta_{m,0}-\delta_{l,0}-\delta_{k,0}}\\ \times
 \big[ &f^{abe} f^{cde} (g_{\mu \sigma} g_{\nu \rho} - g_{\mu \rho}
 g_{\nu \sigma}) +\\ &f^{ace} f^{bde} (g_{\mu \sigma} g_{\nu \rho} -
 g_{\mu \nu} g_{\rho \sigma})+\\ &f^{ade} f^{bce} (g_{\mu \rho} g_{\nu
 \sigma} - g_{\mu \nu} g_{\rho \sigma}) \big]\; ,
\end{split}\\
\end{equation}}\\
\BBSS{a \; \mu}{b \; \nu}{c}{d}{(n) \; k}{(m) \; p}{(l) \; q}{(k) \; r}
&\parbox{130mm}{
\begin{equation}
\begin{split}
\label{quartic}
\Gamma^{abcd}_{\mu \nu 5 5} &(k_{(n)},p_{(m)},q_{(l)},r_{(k)})\ =\ i g^2
\sqrt{2}^{\; -2 -\delta_{n,0}-\delta_{m,0}}\\ &\times
\tilde{\Delta}_{n,m,l,k} \, g_{\mu \nu} \big[ f^{ace} f^{bde} +
f^{ade} f^{bce} \big]\; ,
\end{split}
\end{equation}}
\end{align*}
\begin{align*}
\GGBeast{a}{b}{c \; \mu}{(n) \; k}{(m) \; p}{(l) \; q}
&\parbox{130mm}{
\begin{equation}
\begin{split}
\Gamma^{abc}_{\overline{c} c \mu} &(k_{(n)}, p_{(m)}, q_{(l)})\ =\ \\ &-g
f^{abc} \sqrt{2}^{\; -1-\delta_{n,0}-\delta_{m,0}-\delta_{l,0}} \,
\Delta_{n,m,l} \, k^{\mu}\; ,
\end{split}
\end{equation}}\\ &\phantom{}\\
\GGSeast{a}{b}{c}{(n) \; k}{(m) \; p}{(l) \; q}
&\parbox{130mm}{
\begin{equation}
\begin{split}
\label{ghostScalar}
\Gamma^{abc}_{\overline{c} c 5} &(k_{(n)}, p_{(m)}, q_{(l)})\ =\ \\ &i g
f^{abc} \xi m_l \sqrt{2}^{\; -1-\delta_{n,0}-\delta_{m,0}} \,
\Delta_{n,m,l}\; .
\end{split}
\end{equation}}
\end{align*}
\vspace{1cm}


\setcounter{equation}{0}
\section{Mass Eigenmode Wavefunctions}
\label{masseigenmodeexpansion}

Our aim here is  to determine  the  analytic forms of the  orthonormal
wavefunctions   $f_n   (y)$  and   $g_n     (y)$,   which  are    used
in~(\ref{AmuExpansion}) to express  the 5D fields  $A^a_\mu (x,y)$ and
$A^a_5  (x,y)$ in terms  of the KK  mass eigenmodes $A^a_{(n)\mu} (x)$
and $A^a_{(n)5}  (x)$. In addition,  we will derive the transcendental
equation~(\ref{spectrum}) that determines  the masses for the KK gauge
fields.

We start our discussion  with the observation that the gauge-invariant
part of the 5D QCD Lagrangian~(\ref{LagYMBKT}) is proportional to $[ 1
+ r_c \delta (y) ]$. Then, our approach to quantization in the $R_\xi$
gauge  can be  consistently formulated   if this $\delta(y)$-dependent
factor is promoted to an overall integral weight of the full quantized
action that  includes the gauge-fixing  and Faddeev--Popov ghost terms
given in~(\ref{gfTerm}).  In  such  an  approach, it is  important  to
require that  total derivatives  of  fields or product  of fields with
respect to $y$   do  not   contribute to    the action, but     vanish
identically. This  condition on the surface  terms  can be achieved by
requiring that
\begin{equation}
  \label{surfcond}
\int_{-\pi R+\epsilon}^{\pi R +\epsilon} dy\, \big[  1 +   r_c  \delta
(y) \big]\; \partial_5 f_n (y)\ =\ 0\,,\qquad
\int_{-\pi R+\epsilon}^{\pi R + \epsilon} dy\, \big[  1 +   r_c  \delta
(y) \big]\; \partial_5 g_n (y)\ =\ 0\, . 
\end{equation}
Bear in  mind that our  compact space $y$  is defined in  the interval
$(-\pi  R +\epsilon,\  \pi R  + \epsilon  ]$, where  $\epsilon$  is an
infinitesimal positive constant  which is taken to zero  at the end of
the   calculation.   Using  the   convolution  methods   presented  in
Appendix~C,  it is  not  difficult to  show  that (\ref{surfcond})  is
sufficient  to  ensure the  vanishing  of  any  total derivative  when
integrated along  with the  $\delta (y)$-dependent weight  $[ 1  + r_c
\delta (y) ]$.  As a  consequence of this, the undesirable mixing term
$(\partial_\mu A_5)\,(\partial_5 A^\mu )$ in the gauge-kinetic part of
the 5D  Lagrangian cancels  against a corresponding  term $(\partial_5
A_5)\,   (\partial_\mu  A^\mu)$  that   occurs  in   the  gauge-fixing
Lagrangian~(\ref{gfTerm}).

In addition to~(\ref{surfcond}), the wavefunctions  $f_n (y)$ and $g_n
(y)$ have to satisfy the orthonormality conditions:
\begin{equation}
  \label{ortho}
\begin{split}
\int_{-\pi R+\epsilon}^{\pi R+\epsilon} dy \; 
\big[\, 1\: +\: r_c \delta(y)\,\big] &\,
f_k(y) f_l(y)\ =\ \delta_{k,l}\; ,\\[3mm]
\int_{-\pi R+\epsilon}^{\pi R+\epsilon} dy \; 
\big[\, 1\: +\: r_c \delta(y)\,\big] &\,
g_k(y) g_l(y)\ =\ \delta_{k,l}\; .
\end{split}
\end{equation}
The analytic form of $f_n (y)$ and $g_n (y)$ can be fully specified by
requiring that the kinetic part of the effective 4D Lagrangian for the
KK modes $A^{a}_{(n)\mu}$ and $A^a_{(n) 5}$ takes on the expected form
in the conventional $R_\xi$ gauge:
\begin{eqnarray}
  \label{4DRxi} 
{\cal L}^{\rm eff}_{\rm kin} & = & -\, \frac{1}{4}\;
\big( \partial_\mu A^{a}_{(n)\,\nu} - \partial_\nu
A^{a}_{(n)\,\mu}\big)\, \big( \partial^\mu A^{a\,\nu}_{(n)} -
\partial^\nu A^{a\,\mu}_{(n)}\big)\: -\: \frac{1}{2\xi}\,
\big(\partial_\mu A^{a\,\mu}_{(n)}\big)^2\: +\: \frac{1}{2}\, m^2_n
A^{a}_{(n)\,\mu}\,A^{a\,\mu}_{(n)} \nonumber\\
&&+\, \frac{1}{2}\, \big(\partial_\mu A^{a}_{(n)\,5}\big)\,
\big(\partial^\mu A^{a}_{(n)\,5}\big)\: -\: \frac{\xi}{2}\, m^2_n\,
\big(A^{a}_{(n)\,5}\big)^2\ .
\end{eqnarray}
The   above form of the  Lagrangian~(\ref{4DRxi})  is obtained if $f_n
(y)$ and $g_n (y)$ satisfy the simple wave equations
\begin{eqnarray}
  \label{fnWaveEquation}
\partial^2_5\, f_n(y)\ +\ m_n^2\, f_n(y) &=& 0\; ,\\
\label{gnWaveEquation}
\partial^2_5\, g_n(y)\ +\ m_n^2\, g_n(y) &=& 0\; .
\end{eqnarray}
To find the  solutions to the above wavefunction  equations, we should
consistently      implement      the      constraints~(\ref{surfcond})
and~(\ref{ortho}). Notice that our  approach is different from the one
presented in the existing literature~\cite{Santiago}.

Let us now describe how to find the analytic form of the wavefunctions
$f_n  (y)$.   Because of the  presence  of  the function  $\delta (y)$
through the constraints, our starting point is the piecewise ansatz
\begin{equation}
  \label{fnsol}
f_n(y)\ =\ 
\begin{cases}
\ A_{\I} \sin m_n y\ +\ B_{\I} \cos m_n y\,, \qquad &\textrm{for}
\quad -\pi R +\epsilon\, <\, y\, \leq\, -\,\epsilon\\ 
\ A_{\II} \sin m_n y\ +\ B_{\II} \cos m_n y\,, &\textrm{for}
\qquad \quad \epsilon\, \leq\, y\, \leq\, \pi R+\epsilon \; .
\end{cases}
\end{equation}
Obviously,  our   regularized   ansatz~(\ref{fnsol})  is   a  solution
to~(\ref{fnWaveEquation}), which  excludes  an infinitesimal  interval
$(-\epsilon, \epsilon)$  in the neighbourhood   of the singular  point
$y=0$.   The difference between  the  wavefunctions $f_n (y)$ and $g_n
(y)$  is that the  $f_n (y)$ is an  even function of $y$, whereas $g_n
(y)$ is an odd one.  Moreover, we must assume that  both $f_n (y)$ and
$g_n (y)$ are  periodic functions of $y$, i.e.~$f_n  (y  ) = f_n  (y +
2\pi R)$ and $g_n  (y) = g_n  (y + 2\pi  R)$.  Within their definition
interval $(-\pi R+\epsilon, \pi R  +\epsilon]$, the periodic condition
for the points $y = \pm \pi R$ amounts to the constraints:
\begin{equation}
  \label{periodic}
f_n (-\pi R + \epsilon)\ =\ f_n (\pi R +\epsilon)\,,\qquad 
g_n (-\pi R + \epsilon)\ =\ g_n (\pi R +\epsilon)\,. 
\end{equation}
Although the above periodicity constraint  on $f_n (y)$ may be trivial
because  $f_n (y)$ is  an even  function, the  one applied  to $y$-odd
functions, such  as $\partial_5  f_n (y)$ and  $g_n (y)$,  can provide
useful information (see discussion below).

\begin{figure}[h!]
\begin{center}
\parbox{120mm}{
\begin{center}
\includegraphics[width=0.6\textwidth]{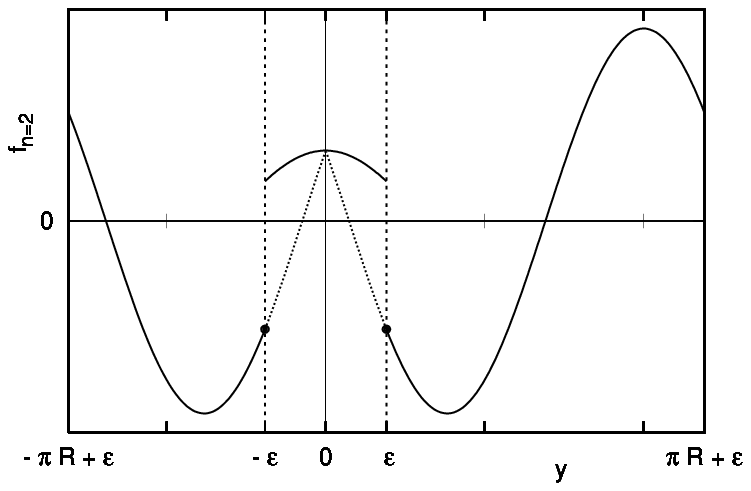}\\
\includegraphics[width=0.6\textwidth]{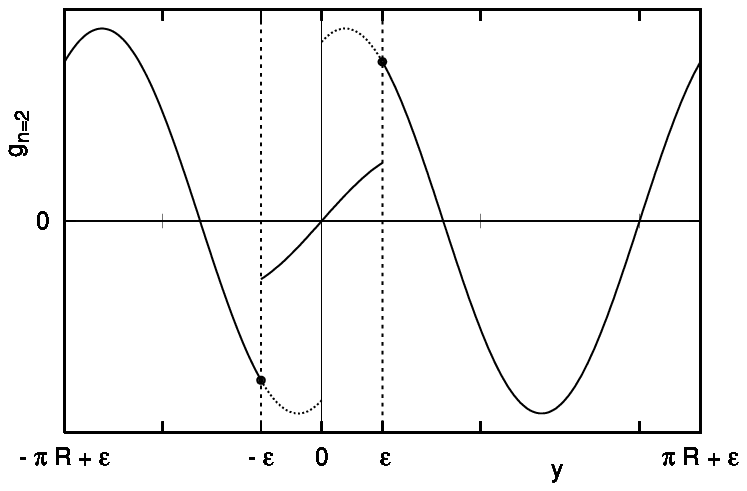}
\vspace{-1cm}
\end{center}}
\end{center}
\caption{\em The  $y$-profile   of   the $\epsilon$-regularized   mass
eigenmode wavefunctions $f_{n=2}  (y)$  and $g_{n=2} (y)$.  The dotted
lines  indicate the $y$-dependence of the   wavefunctions in the limit
$\epsilon \to 0$,  in  the neighbourhood of  the singular  point $y=0$
[cf.\ (\ref{fn}) and (\ref{gn})].}
\label{interpol}
\end{figure}

The unknown coefficients $A_{\rm I,II}$ and $B_{\rm I,II}$, as well as
the  KK  masses $m_n$,   may  be determined    as follows.  {}From the
requirement that $f_n (y)$  is an even  function, i.e.~$f_n (y) =  f_n
(-y)$, we get the constraint
\begin{equation}
  \label{AB}
A_{\I}\ =\ - A_{\II}\ =\ A_n\,,\qquad B_{\I}\ =\ B_{\II}\ =\ B_n\; .
\end{equation}
Imposing the     first condition of~(\ref{surfcond})   on  the $y$-odd
expression $\partial_5 f_n  (y)$ leads to  the constraint: 
\begin{equation}
  \label{formal1}
 \partial_5 f_n (y=0)\ =\ 0\; .
\end{equation}
To     implement~(\ref{formal1})    in  agreement      with  the  wave
equation~(\ref{fnWaveEquation})  for $f_n (y)$, we analytically define
$f_n(y)$ in the interval $-\epsilon < y < \epsilon$ as follows:
\begin{equation}
  \label{interfn}
f_n (y) \ =\ R_n\, \cos m_n y\; .
\end{equation}
The arbitrary constant $R_n$ may be determined by the matching
condition
\begin{equation}
  \label{matching1}
R_n\  =\ \lim_{\epsilon\to 0}\; 
\bigg(\, \lim_{y \to  \epsilon^+}  f_n (y)\,\bigg)\ =\
\lim_{\epsilon\to 0}\; \bigg(\, \lim_{y \to -\epsilon^-} f_n (y)\,\bigg)\; .
\end{equation}
Notice that the above double limits are well-defined, even though $f_n
(y)$ has finite jumps at the  critical points $y = \pm \epsilon$.  The
derivative $\partial_5 f_n (y)$ of the $\epsilon$-regularized function
$f_n (y)$ is also analytically well-behaved at the singular point $y =
0$,    where   $\delta    (y)$   acts,    and   complies    with   the
constraint~(\ref{formal1}).   In  Fig.~\ref{interpol}  we display  the
$y$-profile of the  $\epsilon$-regularized function $f_{n=2} (y)$, and
show its shape in the limit  $\epsilon \to 0$.  Here, we should stress
again  that our  regularization method  consists in  taking  the limit
$\epsilon \to 0$ at the very end of our calculation.

As we will  see below, $\partial_5 f_n  (y)$  is an odd  function that
shares the same properties  as $g_n (y)$.  We may understand this fact
by observing that if the even function $f_n (y)$ is  a solution to the
wave equation~(\ref{fnWaveEquation}),   its derivative $\partial_5 f_n
(y)$,   which        is   an  odd      function,       is  a  solution
to~(\ref{fnWaveEquation})   as     well.  Thus,    assuming  {\it  \`a
posteriori} that,  up to a constant, the  odd function $\partial_5 f_n
(y)$  equals   $g_n   (y)$,  we  may   impose    the second  condition
of~(\ref{surfcond}) in the form
\begin{equation}
  \label{2ndcond}
\int_{-\pi R+\epsilon}^{\pi R+\epsilon} dy \; 
\big[\, 1\: +\: r_c \delta(y)\,\big]\; \partial_5\, 
\big(\partial_5 f_n (y)\big)\ =\ 0\; .
\end{equation}
Using the wave equation~(\ref{fnWaveEquation}) for the function $f_n$,
we get the new constraint:
\begin{equation}
  \label{3rdcond}
\partial_5 f_n (\pi R +\epsilon)\: -\: \partial_5 f_n (-\pi R +\epsilon)\: 
-\: \partial_5 f_n (\epsilon )\: +\: \partial_5 f_n (-\epsilon )\ =\
m_n^2\, r_c\, f_n (0)\; .  
\end{equation}
In     deriving~(\ref{3rdcond}),      we    have     neglected   terms
$\ord{(\epsilon)}$   that  occur in  the     region $-\epsilon  < y  <
\epsilon$.

Our ultimate constraint arises from the fact that  in addition to $f_n
(y)$ which is periodic by  construction, $\partial_5 f_n (y)$ has also
to be periodic on the entire $y$ interval, i.e.~$\partial_5 f_n (y ) =
\partial_5 f_n (y + 2\pi R)$. Applying the periodicity restriction for
the points  $y =  \mp\pi R +\epsilon$  that  lie within the definition
interval, we get the non-trivial relation
\begin{equation}
  \label{periodic2}
\partial_5\, f_n (-\pi R + \epsilon) \ =\ \partial_5\, f_n (\pi R +
\epsilon )\; . 
\end{equation}
The  constraints~(\ref{3rdcond}) and    (\ref{periodic2})  lead to   a
determination of the  coefficients $A_n$ and $B_n$  and the KK  masses
$m_n$:
\begin{equation}
  \label{AnBn}
A_n\ =\ \frac{m_n r_c}{2}\ B_n\; ,\qquad \sin m_n \pi R\: +\: 
\frac{m_n r_c}{2}\, \cos m_n \pi R\ =\ 0\; . 
\end{equation}
The   second   equation   in~(\ref{AnBn})   is   equivalent   to   the
transcendental equation~(\ref{spectrum}).  The remaining freedom is an
overall normalization  constant $N_n$  given in~(\ref{fn}) and  can be
determined from the orthonormality condition~(\ref{ortho}).

To   determine  the analytic  form  of   $g_n  (y)$, we  proceed  very
analogously. We start with a similar ansatz
\begin{equation}
  \label{gnsol}
g_n(y)\ =\
\begin{cases}
\ C_{\I} \sin m_n y\: +\: D_{\I} \cos m_n y\;, \qquad &\textrm{for}
\quad -\pi R +\epsilon\, <\, y \, \leq\, -\epsilon\\ 
\ C_{\II} \sin m_n y\: +\: D_{\II} \cos m_n y\;, &\textrm{for}
\qquad \quad \epsilon\, \leq\, y\, \leq\, \pi R +\epsilon\; .
\end{cases}
\end{equation}
{}From the fact that $g_n (y)$ is an odd function, we get
\begin{equation}
  \label{CD}
C_{\I}\ =\ C_{\II}\ =\ C_n\,,\qquad D_{\I}\ =\ - D_{\II}\ =\ D_n\; .
\end{equation}
In  view of    the constraints~(\ref{CD}),   $g_n    (y)$ exhibits   a
discontinuity in the limit $y=\pm \epsilon \to  0$.  Exactly as we did
above for $f_n   (y)$,  we analytically  regularize $g_n  (y)$  in the
interval $(-\epsilon, \epsilon)$ as
\begin{equation}
  \label{intergn}
g_n (y)\ =\ Q_n\, \sin m_n y\; .  
\end{equation}
Observe  that  (\ref{intergn})   is  also  a  solution   to  the  wave
equation~(\ref{gnWaveEquation})  for the   function  $g_n (y)$.  Since
$\partial_5 g_n (y)$  determined from~(\ref{gnsol})  has  well-defined
limits when $y=\pm \epsilon \to 0$, we apply the matching condition
\begin{equation}
  \label{matching2}
m_n\,Q_n\  =\ \lim_{\epsilon\to 0}\; 
\bigg(\, \lim_{y \to  \epsilon^+}\, \partial_5 g_n (y)\,\bigg)\ =\
\lim_{\epsilon\to 0}\; \bigg(\, \lim_{y \to -\epsilon^-}\, 
\partial_5 g_n (y)\,\bigg)\;  
\end{equation}
to determine the   constant $Q_n$.  Given~(\ref{intergn}), the  second
equation of the condition~(\ref{surfcond}) implies that
\begin{equation}
  \label{4thcond}
g_n (\pi R +\epsilon)\: -\: g_n (-\pi R +\epsilon)\: 
-\: g_n (\epsilon )\: +\: g_n (-\epsilon )\ =\ 
-\,r_c\, \partial_5 g_n (y=0)\; .  
\end{equation}
In writing down~(\ref{4thcond}),  we also made   use of the  fact that
terms $\ord{(\epsilon )}$ arising from the integration in the interval
$(-\epsilon, \epsilon)$  have been  neglected.  For  illustration,  we
show  in     Fig.~\ref{interpol}     the   $y$-dependence   of     the
$\epsilon$-regularized  wavefunction $g_{n=2}   (y)$,  along with  its
$y$-profile in the limit $\epsilon \to 0$.

Finally, taking  into  account the periodic condition~(\ref{periodic})
for the functions $g_n (y)$, we obtain the relation
\begin{equation}
D_n\ =\ -\; \frac{ m_n r_c}{2}\ C_n\; ,
\end{equation}
and the  second   equation in~(\ref{AnBn}).   As before,   the overall
normalization constant  can  be  determined  from the   orthonormality
condition~(\ref{ortho}).   Finally,   implementing     the    matching
conditions (\ref{matching1})  and (\ref{matching2}), we find that $R_n
= Q_n$.

After having  derived the analytic  forms of  $f_n (y)$ and  $g_n (y)$
given in (\ref{fn}) and  (\ref{gn}), respectively, it is not difficult
to  see   that $f_n    (y)$  and $g_n   (y)$  satisfy   the  relations
in~(\ref{partial}),  for the entire  piecewise defined region $(-\pi R
+\epsilon, \pi R + \epsilon  ]$ that also  includes the singular point
$y=0$.   Finally, we should  remark  that  for  a 5D  orbifold  theory
without BKTs, the functions $f_n (y)$ and $g_n (y)$ go to the standard
Fourier expansion modes in limit $r_c \to 0$:
\begin{equation}
\label{lim}
\lim_{r_c \to 0}\, f_n(y)\ =\ \frac{1}{\sqrt{ 2^{\delta_{n,0}} \pi R}} \;
\cos \frac{ny}{R}\ , \qquad  \lim_{r_c \to 0}\, g_n(y)\ =\
\frac{1}{\sqrt{\pi R}} \; \sin \frac{ny}{R}\ .
\end{equation}

\begin{figure}[h!]
\begin{center}
\parbox{120mm}{
\begin{center}
\includegraphics[width=0.7\textwidth]{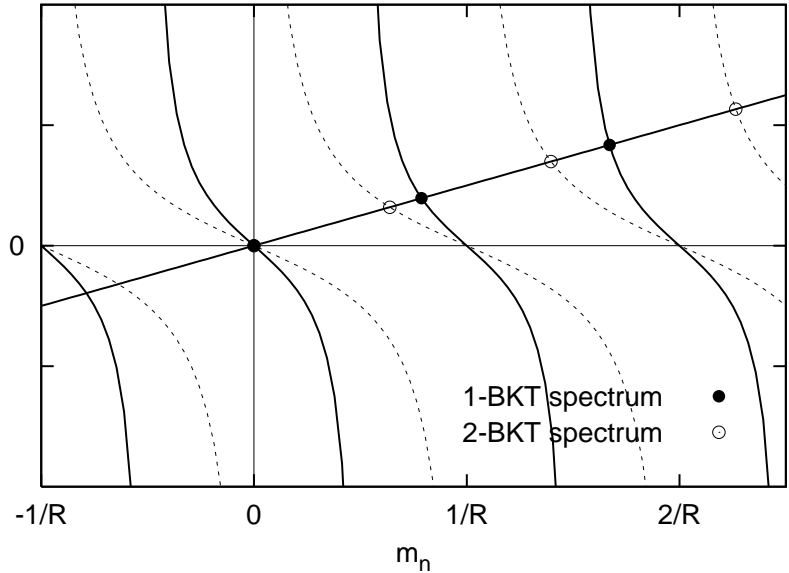}
\end{center}}
\end{center}
\caption{{\em Mass spectrum $m_n$ of the effective 4D theory with
one/two BKT in the 5D Lagrangian, (\ref{spectrum}) and
(\ref{spectrum2BKT}), respectively.}} \label{spectra}
\end{figure}

For   completeness,   we present in  the  remainder   of this appendix
analytic results concerning the case of  a 5D orbifold theory with two
BKTs at the orbifold fixed points $y = 0$ and $y  = \pi R$.  We follow
a very analogous approach to the one given above for the one BKT case.
We start   again    with the   same   form  of   regularized  ansaetze
(\ref{fnsol}) and (\ref{gnsol}) for $f_n (y)$ and $g_n (y)$, which are
piecewise defined  in the  interval $(-\pi  R  + \epsilon, -\epsilon]\
\cup\ [\epsilon  , \pi R  -\epsilon]$.  Notice that the $y$-intervals,
$(-\epsilon,\epsilon)$ and $(\pi R -\epsilon, \pi R + \epsilon ]$, are
excluded, because of the singular overall integral measure: $\big[\, 1
+ r_c\,\delta (y) +  r_c \delta (y -  \pi R)\,\big]$. However, most of
the   discussion given   above   goes  through,  with    some  obvious
modifications related to the presence of the second $\delta$-function,
$\delta (y  - \pi R)$.  In  fact, the only  crucial difference is that
the constraints~(\ref{3rdcond}) and~(\ref{4thcond}) now become
\begin{eqnarray}
  \label{2BKTcond1}
\partial_5 f_n (\pi R -\epsilon)\: -\: \partial_5 f_n (-\pi R +\epsilon)\: 
-\: \partial_5 f_n (\epsilon )\: +\: \partial_5 f_n (-\epsilon )& =&
m_n^2\, r_c\, \big[\, f_n (0)\: +\: f_n (\pi R)\,\big]\; ,\nonumber\\
&& \\[3mm]  
  \label{2BKTcond2}
g_n (\pi R -\epsilon)\: -\: g_n (-\pi R +\epsilon)\: 
-\: g_n (\epsilon )\: +\: g_n (-\epsilon ) &=&
-\,r_c\, \big[\, \partial_5 g_n (0)\: +\: \partial_5 g_n (\pi
R)\,\big]\; .\nonumber\\
&&   
\end{eqnarray}
Taking into consideration the above non-trivial modifications, we find
that the  mass eigenmode wavefunctions  $f_n(y)$ and $g_n  (y)$ retain
the functional form  given in (\ref{fn}) and (\ref{gn}),  in the limit
$\epsilon \to  0$, for the  one BKT case.  However,  the normalization
factor $N_n$ modifies in the presence of the second BKT at $y = \pi R$. 
The analytic form of $N_n$ may be determined by
\begin{equation}
\label{Nn2BKT}
\begin{split}
N_n^{-2}\ =& \ 1\: +\: 4\, \tilde{r}_c\: 
+\: \frac{1+2 \tilde{r}_c}{(1-\tilde{r}_c \pi R m_n)^2}\: 
+\: \frac{1+2 \tilde{r}_c}{(1+\tilde{r}_c \pi R m_n)^2}\\
&-\ \frac{1+3\tilde{r}_c}{1-\tilde{r}_c \pi R m_n}\:
-\: \frac{1+3\tilde{r}_c}{1+\tilde{r}_c \pi R m_n}\ .
\end{split}
\end{equation}
Moreover, the transcendental equation determining the mass spectrum of
the effective 4D theory may be cast into the useful factorizable form:
\begin{equation}
  \label{spectrum2BKT}
m_n\, \bigg[\, \tan \bigg(\,\frac{m_n \pi R}{2}\,\bigg)\ +\ 
\frac{m_n r_c}{2}\, \bigg]\, 
\bigg[\, \cot \bigg(\,\frac{m_n \pi R}{2}\,\bigg)\ -\ 
\frac{m_n r_c}{2}\, \bigg]\ =\ 0\; ,
\end{equation}
which  is valid for  $r_c  \neq 0$   and,  after some  algebra, agrees
with~\cite{Carena}.   In    Fig.~\ref{spectra},  we  present   typical
solutions to   (\ref{spectrum})  and  (\ref{spectrum2BKT})  for   a 5D
orbifold theory with one and two BKTs, respectively.

\newpage
\section{Convolution Integrals for $S^1/\Zb_2$ Wavefunctions}
\label{multiplication}
\setcounter{equation}{0}

In the process of the KK reduction,  expressions that involve products
of 5D fields compactified on an $S^1/\Zb_2$ orbifold, such as $A^a_\mu
(x,y)$ and $A^a_5  (x,y)$,  need to be   integrated over  the  compact
dimension  $y$.  These 5D  fields have  an   even or odd parity  under
$\Zb_2$ transformations  and can be expressed in  terms of the KK mass
eigenmodes $f_n  (y)$  and    $g_n  (y)$,  given in   (\ref{fn})   and
(\ref{gn}), respectively.   Using convolution integral  techniques, we
may express the product of two $\Zb_2$-even functions, e.g.~$F(y)$ and
$G(y)$,  in a series of  the   orthonormal wavefunctions $f_n (y)$  as
follows:
\begin{equation}
F(y)\,G(y)\ =\ \sum_{n=0}^{\infty}\, [F*G]_{(n)}\, f_n(y) \, ,
\end{equation}
where   the coefficients $[F*G]_{(n)}$ are   given  by the convolution
integrals
\begin{equation}
  \label{FGn}
[F*G]_{(n)}\ =\ \sum_{k,l=0}^{\infty} F_{(k)} G_{(l)} 
\int_{-\pi R}^{\pi R} dy \, \big[ 1 + r_c \delta(y) \big]\, 
f_k(y) f_l(y) f_n(y)\; .
\end{equation}
Likewise,  the product  of  a  $\Zb_2$-even  function  $F(y)$ with   a
$\Zb_2$-odd one $H(y)$ can be expanded in  a series of the orthonormal
wavefunctions $g_n (y)$:
\begin{equation}
F(y)\,H(y)\ =\ \sum_{n=0}^{\infty}\, [F*H]_{(n)}\, g_n(y) \, ,
\end{equation}
where the  coefficients  $[F*H]_{(n)}$  are given by   the convolution
integrals
\begin{equation}
  \label{FHn}
[F*G]_{(n)}\ =\ \sum_{k,l=0}^{\infty} F_{(k)} H_{(l)} 
\int_{-\pi R}^{\pi R} dy \, \big[ 1 + r_c \delta(y) \big]\, 
f_k(y) g_l(y) g_n(y)\; .
\end{equation}

The integrals~(\ref{FGn}) and (\ref{FHn}) appear in the calculation of
the   effective Feynman rules,   and  of the effective   gauge and BRS
transformations, when the 5D fields are expressed in terms of their KK
modes. Specifically, defining the overlap integrals as
\begin{equation}
\begin{split}
\int_{-\pi R}^{\pi R} dy \, \big[ 1 + r_c \delta(y) \big] \, 
f_k(y) \, f_l(y) \, f_n(y)
\ =&\ \frac{\Delta_{k,l,n}}{2 
\sqrt{2^{\delta_{k,0}+\delta_{l,0}+\delta_{n,0}} \pi R}} \ , \\
\int_{-\pi R}^{\pi R} dy \, \big[ 1 + r_c \delta(y) \big] \, 
f_k(y) \, g_l(y) \, f_n(y)
\ = &\ \frac{\tilde{\Delta}_{k,l,n}}{2 
\sqrt{2^{\delta_{l,0}} \pi R}} \ ,
\end{split}
\end{equation}
we may  analytically  calculate the coefficients  $\Delta_{k,l,n}$ and
$\tilde{\Delta}_{k,l,n}$ as follows:
\begin{equation}
\label{DeltaDownThree}
\begin{split}
\Delta_{k,l,n}\ & = \ 
\Delta(m_k, m_l, m_n)\: +\: \Delta(-m_k, m_l, m_n)\:
+\: \Delta(m_k, -m_l, m_n)\: +\: \Delta(m_k, m_l, -m_n) \, ,\\[1mm]
\tilde{\Delta}_{k,n,l}\ &=\ -\,\Delta(m_k, m_l, m_n) \:
-\: \Delta(m_k, m_l, -m_n)\: +\: \Delta(-m_k, m_l, m_n) \:
+\: \Delta(m_k, -m_l, m_n) \, ,
\end{split}
\end{equation}
where
\begin{equation}
\label{DeltaThree}
\Delta(m_k,m_l,m_n)\ =\ N_k N_l N_n\, \left[\, \frac{\int_0^{\pi R} dy \, 
\cos (m_k+m_l+m_n)(y-\pi R)}{\pi R \, \cos m_k \pi R \, \cos m_l \pi R \, 
\cos m_n \pi R}\ +\ \tilde{r}_c\, \right] \, .
\end{equation}
The  normalization   constants  $N_n$   are given in   (\ref{Nn})  and
$\tilde{r}_c = r_c/(2 \pi R)$.  According to the above definition, the
coefficient $\Delta_{k,l,n}$ is symmetric in all of its indices, while
$\tilde{\Delta}_{k,l,n}$  is only  symmetric under the  interchange of
the first and the last index.  Note that the argument of the cosine in
(\ref{DeltaThree})  may   vanish,   when some  of    the arguments  in
$\Delta(m_k,m_l,m_n)$ become negative.   Taking  this possibility into
account, we find
\begin{align}
  \label{Dmasskln}
\Delta(m_k,m_l,m_n)\
&= \
\begin{cases}
& \hspace{-3mm} N_k N_l N_n\, \pi^2 R^2 \tilde{r}_c^3 \; \frac{\displaystyle
m_k m_l m_n}{\displaystyle m_k+m_l+m_n}\ ,
\qquad \qquad 
\textrm{for} \quad m_k+m_l+m_n \neq 0 \\[2ex]
\begin{split}
& \hspace{-3mm} N_k N_l N_n \Big\{
\Big[(1+\pi^2 R^2 \tilde{r}_c^2 m_k^2)(1+\pi^2 R^2 
\tilde{r}_c^2 m_l^2)(1+\pi^2 R^2 \tilde{r}_c^2
m_n^2)\Big]^{\frac{1}{2}}\: +\:
\tilde{r}_c \Big\},\\
& \qquad \qquad \qquad \qquad \qquad \qquad \qquad \qquad \qquad \,
\textrm{for} \quad m_k+m_l+m_n = 0\; .
\end{split}
\end{cases}
\end{align}
For non-vanishing $\tilde{r}_c$, the  lower case is only important for
the  calculation  of $\Delta(m_k,m_l,m_n)$, if    at least one of  the
indices $k$, $l$, $n$ is zero and the other two  are equal. When this
condition is satisfied, we find
\begin{eqnarray}
\label{Dnn0}
\Delta_{0,0,0} &=& \frac{4}{\sqrt{1+\tilde{r}_c}}\ , 
\qquad \tilde{\Delta}_{n,n,0}\ =\ 0\; ,\nonumber\\
\Delta_{n,n,0} &=& \tilde{\Delta}_{n,0,n}\ =\ 
\frac{2 N_n^2}{\sqrt{1+\tilde{r}_c}}\,\Big(\, 1\: +\: \tilde{r}_c\: 
+\: \pi^2 R^2 \tilde{r}_c^2 m_n^2\, \Big)\; .
\end{eqnarray}
In    any   other case  (up       to  symmetries), the    coefficients
$\Delta_{k,l,n}$ and $\tilde{\Delta}_{k,l,n}$  have been calculated to
have the more compact analytic forms:
\begin{equation}
\begin{split}
\Delta_{k, l, n}\ & =\ N_k N_l N_n \pi^2 R^2 \tilde{r}_c^3 \; 
\frac{8 m_k^2 m_l^2 m_n^2}{m_k^4+m_l^4+m_n^4-2(m_k^2 m_l^2 + 
m_l^2 m_n^2 + m_n^2 m_k^2)} \ , \\[2mm]
\tilde{\Delta}_{k, n, l}\ & =\ N_k N_l N_n \pi^2 R^2 \tilde{r}_c^3 \;
\frac{4 m_k m_l m_n^2(m_k^2+m_l^2-m_n^2)}{m_k^4+m_l^4+m_n^4-2(m_k^2 
m_l^2 + m_l^2 m_n^2 + m_n^2 m_k^2)} \ .
\end{split}
\end{equation}
If two   of  the indices  are  equal, the  last  two formulae simplify
further to
\begin{eqnarray}
  \label{Dnnj}
\Delta_{n,n,j} &=& N_n^2 N_j \pi^2 R^2 \tilde{r}_c^3 \; 
\frac{8 m_n^4}{m_j^2-4 m_n^2} \ , \nonumber\\
\tilde{\Delta}_{n,n,j} &=& N_n^2 N_j \pi^2 R^2 \tilde{r}_c^3 \; 
\frac{4 m_j m_n^3}{m_j^2-4m_n^2} \ , \\
\tilde{\Delta}_{n,j,n} &=& -\, N_n^2 N_j \pi^2 R^2 \tilde{r}_c^3 
\; \frac{4 m_n^2\, (m_j^2-2m_n^2)}{m_j^2-4 m_n^2} \ .\nonumber
\end{eqnarray}

Correspondingly, we may now   evaluate overlap integrals  that involve
the product of 4 orthonormal wavefunctions,
\begin{equation}
\begin{split}
\int_{-\pi R}^{\pi R} dy \, \big[ 1+r_c \delta(y) \big]\, 
f_k(y) f_l(y) f_n(y) f_m(y)\
& =\ \frac{\Delta_{k,l,n,m}}{4 \pi R
\sqrt{2^{\delta_{k,0}+\delta_{l,0}+\delta_{n,0}+\delta_{n,0}}}} \ ,\\[1ex]
\int_{-\pi R}^{\pi R} dy \, \big[ 1+r_c \delta(y)\big]\, 
f_k(y) f_l(y) g_n(y) g_m(y)
\ & =\ \frac{\tilde{\Delta}_{k,l,n,m}}{4 \pi R 
\sqrt{2^{\delta_{k,0}+\delta_{l,0}}}} \ .
\end{split}
\end{equation}
The coefficients $\Delta_{k,l,n,m}$ and $\tilde{\Delta}_{k,l,n,m}$ are
given by
\begin{eqnarray}
  \label{Dklnm}
\Delta_{k,l,n,m} &=&  \Delta(m_k,m_l,m_n,m_m)\: +\:
\Delta(-m_k,m_l,m_n,m_m)\: +\: \Delta(m_k,-m_l,m_n,m_m)\nonumber\\
&& +\,\Delta(m_k,m_l,-m_n,m_n)\: +\: \Delta(m_k,m_l,m_n,-m_m)\: +\:
\Delta(-m_k,-m_l,m_n,m_m)\nonumber\\ 
&& +\, \Delta(-m_k,m_l,-m_n,m_m)\: +\:
\Delta(-m_k,m_l,m_n,-m_m) \, , \\[2mm]
  \label{tDklnm}
\tilde{\Delta}_{k,l,n,m} &=& -\,\Delta(m_k,m_l,m_n,m_m)\: -\:
\Delta(-m_k,m_l,m_n,m_m)\: -\: \Delta(m_k,-m_l,m_n,m_m)\nonumber\\ 
&& -\, \Delta(-m_k,-m_l,m_n,m_m)\: +\: \Delta(m_k,m_l,-m_n,m_m)\: +\:
\Delta(m_k,m_l,m_n,-m_m)\nonumber\\ 
&& +\, \Delta(-m_k,m_l,-m_n,m_m)\: +\: \Delta(-m_k,m_l,m_n,-m_m) \, ,
\end{eqnarray}
where     $\Delta(m_k,m_l,m_n,m_m)$   is  an   obvious  generalization
of~(\ref{DeltaThree}), i.e.
\begin{equation}
\label{DeltaFour2}
\begin{split}
\Delta&(m_k,m_l,m_n,m_m)\ =\ N_k N_l N_n N_m\\[2mm]
&\times\, \left\{ 
\begin{array}{l}
 \pi^2 R^2 \tilde{r}_c^3 \ 
\frac{\displaystyle m_l m_n m_m + m_k m_n m_m + m_k m_l m_m + 
m_k m_l m_n}{\displaystyle m_k+m_l+m_n+m_m}\ , \\[1ex]
\qquad \qquad \qquad \qquad \qquad \qquad \qquad \qquad \qquad \qquad
\textrm{for} \quad m_k+m_l+m_n+m_m \neq 0\\[2ex]
\Big\{ \Big[(1+\pi^2 R^2 \tilde{r}_c^2 m_k^2)
(1+\pi^2 R^2 \tilde{r}_c^2 m_l^2)(1+\pi^2 R^2 \tilde{r}_c^2 m_n^2)
(1+\pi^2 R^2 \tilde{r}_c^2 m_m^2) \Big]^{\frac{1}{2}}\: +\: 
\tilde{r}_c \Big\}\,,\\[1ex]
\qquad \qquad \qquad \qquad \qquad \qquad \qquad \qquad \qquad \qquad
\textrm{for} \quad m_k+m_l+m_n+m_m = 0 \, \, .
\end{array}
\right.
\end{split}
\end{equation}
It is easy to  see that $\Delta_{k,l,n,m}$  is symmetric in all of its
indices,   while $\tilde{\Delta}_{k,l,n,m}$  is  only  symmetric in the first pair as well as the last two of its indices.  If  one of the 4
indices is zero  in $\Delta_{k,l,n,m}$ and $\tilde{\Delta}_{k,l,n,m}$,
then (\ref{Dklnm}) and (\ref{tDklnm}) reduce to
\begin{equation}
  \label{Dkln0}
\Delta_{k,l,n,0}\ =\ \frac{2 \Delta_{k,l,n}}{\sqrt{1+ \tilde{r}_c}}\ ,\qquad 
\tilde{\Delta}_{0,k,l,n}\ =\ \frac{2 \tilde{\Delta}_{l,k,n}}{\sqrt{1+ 
\tilde{r}_c}} \ .
\end{equation}
It is also useful to list analytic results for special combinations of
indices, when the lower case in~(\ref{DeltaFour2}) becomes relevant:
\begin{eqnarray}
  \label{Dnnnn}
\Delta_{0,0,0,0}&=&\frac{8}{1+\tilde{r}_c} \ , \nonumber\\[1ex]
\Delta_{n,n,m,m} &=& 2 N_n^2 N_m^2\, \Big[\, 1\: +\: \tilde{r}_c\: +\: 
(1- \tilde{r}_c) \tilde{r}_c^2 \pi^2 R^2 (m_n^2+m_m^2)\: +\: 
\tilde{r}_c^4 \pi^4 R^4 m_n^2 m_m^2\, \Big] \, ,\qquad \\[1ex]
\Delta_{n, n, n, n} &=& 3 \tilde{\Delta}_{n,n,n,n}\ =\ 
3 N_n^4\, \Big[\, \tilde{r}_c(1-\pi^2 R^2 \tilde{r}_c^2 m_n^2)\: +\:
(1+\pi^2 R^2 \tilde{r}_c^2 m_n^2)^2\, \Big] \, .\nonumber 
\end{eqnarray}

Finally,  let us comment  on the limit of vanishing $r_c$ for our
analytic  results.      Taking   this    limit for    the   expression
$\Delta(m_k,m_l,m_n)$, for example, is a non-trivial task.  One should
take care of the fact that although one  may have $m_k+m_l+m_n \neq 0$
for $r_c  \ne 0$, it could  be that $m_k+m_l+m_n  \to  0$ for $r_c \to
0$. In such a case, the denominator that appears in  the first line of
(\ref{Dmasskln}) approaches zero, and one  has to carefully expand the
BKT  corrections to  the KK  masses  in powers of  $r_c$   to obtain a
sensible result. In this way, we find that
\begin{equation}
\label{limD1}
\lim_{r_c \to 0} \; \Delta(m_k,m_l,m_n)\ =\ \delta_{k+l+n,0} \; .
\end{equation}
Analogous considerations for $\Delta(m_k,m_l,m_n,m_m)$ lead to 
\begin{equation}
\label{limD2}
\lim_{r_c \to 0} \; \Delta(m_k,m_l,m_n,m_m)\ =\ \delta_{k+l+n+m,0}\; .
\end{equation}
Given that $\delta_{k+l+n,0}$ and  $\delta_{k+l+n+m,0}$ are the  usual
Kronecker symbols, we have checked that our analytic results perfectly
agree  with those presented   in~\cite{MPR}, within the context of  5D
Yang--Mills orbifold theories without BKTs.


\section{High Energy Unitarity Sum Rules}
\label{summation}
\setcounter{equation}{0}

In  proving  the ET  in   Section~2, we  encounter  infinite sums that
involve products of  the gauge-coupling  coefficients $\Delta_{k,l,n}$
and      $\tilde{\Delta}_{k,l,n}$            defined     in Appendix~C
[cf.~(\ref{DeltaDownThree})]. For instance, we  have to carry  out the
infinite     sum,    $$\sum_{j=0}^{\infty}\;       2^{-\delta_{j,0}}\,
\Delta^2_{n,n,j}\      ,$$ that   occurs      in the   calculation  of
(\ref{gaugeBKT}). Considering~(\ref{Dnnj}),   the problem  reduces  to
finding an expression for the sum,
$$\sum_{j=1}^{\infty}\   \frac{N_j^2}{(m_j^2-4m_n^2)^2}\     .$$   The
calculation of such infinite sums can  be carried out analytically, by
extending   the complex  integration  analysis   techniques  developed
in~\cite{Paes}.

Let  us  briefly describe our   complex integration analysis method by
considering the infinite sum $$\sum_{j=1}^{\infty}\, N_j^2\; ,$$ where
the normalization factors  $N_j$  are given in~(\ref{Nn}).  We   start
with the complex analytic expression
\begin{equation}
  \label{sameRes}
\frac{1}{z\: +\: \tan \pi R z/(\tilde{r}_c \pi R)}\ \approx\ \bigg( 1+
\frac{1}{\tilde{r}_c \cos^2 \pi R z} \bigg)^{-1} \; \frac{1}{z-m_j}\
\equiv\ \frac{\tilde{r}_c\ N(z)}{z-m_j}\ ,
\end{equation}
where we have expanded it about the regions $|z-m_j|<\epsilon $, close
to its poles   $\pm  m_j$. Here  and in  the   following, we use   the
convention that $m_{-j} = -m_j$ and $m_j \ge 0$, for $j\ge 0$.  In the
last equality of~(\ref{sameRes}), we have defined the function $N(z)$,
in  a way  such  that  it  is $N   (m^2_j)  = N^2_j$ at the   residuum
of~(\ref{sameRes}).

Our next step is to use Cauchy's theorem~\cite{Spiegel} to integrate 
the LHS of~(\ref{sameRes}) over the poles $m_j$, i.e.
\begin{equation}
  \label{CT}
\lim_{n \to \infty} \, \oint_{C_n} dz \, \bigg( 
z + \frac{1}{\tilde{r}_c\pi R} \tan \pi R z \bigg)^{-1}\ =\
\lim_{n \to \infty} \, \oint_{C_n} dz\; \tilde{r}_c\;
\sum_{j=-\infty}^{\infty}\; \frac{N (z)}{z-m_j}\ .
\end{equation}
The  contours $C_n$  are the circles  $z_n=(n+1/2)/R  \; e^{i \theta}$
defined in  the complex plane, whose radii  are taken to infinity in a
discrete manner, i.e.~$n\to \infty$.  This ensures that neither $m_j$,
nor  possible  poles of   the complex function  $N(z)$   do lie on the
contour  $C_n$.   The    LHS   of~(\ref{CT}) then becomes    a  simple
integration over  the angle $\theta$ that can  be  performed easily to
give $2\pi i$.  The RHS of~(\ref{CT}) can be calculated using Cauchy's
theorem. Thus, we arrive at the equality
\begin{equation}
  \label{here}
2 \pi i\ =\ 2 \pi i \, \sum_{j=-\infty}^{\infty} \, \tilde{r}_c N(m^2_j)\; ,
\end{equation}
which is equivalent to the desired result
\begin{equation}
   \label{simpleExample}
\sum_{j=1}^{\infty} N_j^2\ =\ \frac{1}{2 \tilde{r}_c(1+\tilde{r}_c)}\ .
\end{equation}

The above method  can be  extended to  more complicated infinite  sums
that include factors, such as $\tilde{\Delta}^2_{n,j,n}$. For example,
a typical infinite sum that arises in such a calculation is
$$\sum_{j=1}^{\infty}\ \frac{N^2_j\,m_j^4}{(m_j^2-4m_n^2)^2}\ .$$   
In  this case, we may  exploit  the analytic structure  of the summing
terms  with respect to  $m_j$, and  consider the analytic continuation
$m_j \to z$ in the complex plane.  Then, we multiply  both the LHS and
the RHS of~(\ref{sameRes}) with the analytic expression
$$ \frac{z^4}{(z^2-4m_n^2)^{2}}\  .$$  
This extra factor leaves the integral on the LHS of~(\ref{CT}) intact,
since  the new factor goes to  1 in the limit  $|z_n| \to \infty$, for
$n\to  \infty$.   To calculate the     RHS of~(\ref{CT}), we  have  to
properly   include the contributions  from  the  two additional double
poles at $z= \pm 2 m_n$. Thus, we finally obtain
\begin{equation}
  \label{Sexample}
\sum_{j=1}^{\infty}\ \frac{N^2_j\,m_j^4}{(m_j^2-4m_n^2)^2}\ =\
\frac{1}{12 \pi^4 R^4 m_n^4 \tilde{r}_c^6 N_n^4}\; \bigg(
\Delta_{n,n,n,n}\: +\: \frac{3}{4}\;X_n\,\bigg)\; ,
\end{equation}
where   $\Delta_{n,n,n,n}$ and   $X_n$  are    given  in~(\ref{Dnnnn})
and~(\ref{Xn}), respectively.

With the help of  the  complex  integration analysis method   outlined
above, the following  list of high energy  unitarity sum rules  can be
derived:
\begin{eqnarray}
  \label{Snnnn1}
\sum_{j=0}^{\infty} 2^{-\delta_{j,0}} \Delta_{n,n,j}^2 \!&=&\! 
\Delta_{n,n,n,n}\;,\qquad\qquad
\sum_{j=1}^{\infty} \tilde{\Delta}_{n,n,j}^2\ =\
\tilde{\Delta}_{n,n,n,n}\; ,\\
  \label{Snnnn2}
\sum_{j=0}^{\infty} 2^{-\delta_{j,0}} 
\Delta_{n,n,j} \tilde{\Delta}_{n,j,n} \!&=&\! \tilde{\Delta}_{n,n,n,n}\;,
\qquad
\sum_{j=0}^{\infty} 2^{-\delta_{j,0}} \tilde{\Delta}_{n,j,n}^2\ =\
\Delta_{n,n,n,n}\: +\: X_n\; ,\\
  \label{Snnmm1}
\sum_{j=0}^{\infty} 2^{-\delta_{j,0}}
\Delta_{n,j,m}^2 \! &=&\! \Delta_{n,n,m,m}\; ,\quad
\sum_{j=0}^{\infty} 2^{-\delta_{j,0}} \tilde{\Delta}_{n,j,n}
\tilde{\Delta}_{m,j,m} \ =\ \Delta_{n,n,m,m}\: +\: Y_{n,m}\; ,\\
  \label{Snnmm2}
\sum_{j=0}^{\infty} 2^{-\delta_{j,0}} \tilde{\Delta}_{n,j,m}^2\!&=&\!
\Delta_{n,n,m,m}\: +\: Y_{n,m}\; ,\qquad \\ 
  \label{Sklnm}
\sum_{j=0}^{\infty} 2^{-\delta_{j,0}} \tilde{\Delta}_{k,j,l}
\tilde{\Delta}_{n,j,m} \!&=&\! \Delta_{k,l,n,m}\: +\: Z_{k,l,n,m}\; ,
\end{eqnarray}
where
\begin{eqnarray}
  \label{Ynm}
Y_{n,m} &=& 4\,N_n^2 N_m^2\, \pi^2 R^2 \tilde{r}_c^3\, (m_n^2 + m_m^2)\;,\\
  \label{Zklnm}
Z_{k,l,n,m} &=& 2\, N_k N_l N_n N_m\,
 \pi^2 R^2 \tilde{r}_c^3\, (m_k^2 + m_l^2 + m_n^2 +
m_m^2) \; .
\end{eqnarray}
The  sum   rules~(\ref{Snnnn1})  and  (\ref{Snnnn2})  are  relevant to
restore unitarity in high energy $2\to 2$ processes where all KK gauge
modes  in the initial and final  state have the  same mass $m_n$.  The
sum rules~(\ref{Snnmm1})    and  (\ref{Snnmm2})   ensure   high energy
unitarity for $2\to  2$ scatterings, where  the KK gauge modes  in the
initial and  final  states   have  the same  mass   $m_n$ and   $m_m$,
respectively, but $m_n \neq  m_m$.  Finally, (\ref{Sklnm}) is relevant
to  restore unitarity in $2\to 2$  scatterings,  when all 4 asymptotic
states happen to have different masses.

\end{appendix}


\end{document}